\documentclass[bibyear]{aa}  
\usepackage{natbib}
\usepackage{graphicx}
\usepackage{dcolumn}

\usepackage{txfonts}
\usepackage{subcaption}
\usepackage{caption}
\usepackage{color}
\usepackage{natbib,twoopt}
\usepackage[hyphenbreaks]{breakurl}
\usepackage[breaklinks]{hyperref}      
\bibpunct{(}{)}{;}{a}{}{,}             
\definecolor{cobalt}{rgb}{0.06, 0.2, 0.65}
\hypersetup{
  colorlinks,
  citecolor=cobalt,
  linkcolor=[rgb]{0.8, 0.2, 1.0},
  urlcolor=cobalt,}
\makeatletter
  \newcommandtwoopt{\citeads}[3][][]{\href{http://adsabs.harvard.edu/abs/#3}%
    {\def\hyper@linkstart##1##2{}%
     \let\hyper@linkend\@empty\citealp[#1][#2]{#3}}}
  \newcommandtwoopt{\citepads}[3][][]{\href{http://adsabs.harvard.edu/abs/#3}%
    {\def\hyper@linkstart##1##2{}%
     \let\hyper@linkend\@empty\citep[#1][#2]{#3}}}
  \newcommandtwoopt{\citetads}[3][][]{\href{http://adsabs.harvard.edu/abs/#3}%
    {\def\hyper@linkstart##1##2{}%
     \let\hyper@linkend\@empty\citet[#1][#2]{#3}}}
  \newcommandtwoopt{\citeyearads}[3][][]%
    {\href{http://adsabs.harvard.edu/abs/#3}
    {\def\hyper@linkstart##1##2{}%
     \let\hyper@linkend\@empty\citeyear[#1][#2]{#3}}}
\makeatother
\usepackage{siunitx}

\def\ts     {\thinspace} 
\usepackage{amsmath}

\def\kms  {\ifmmode{{\rm \ts km\ts s}^{-1}}\else{\ts km\ts s$^{-1}$\ts}\fi}
\def\msol {\ifmmode{{\rm M}_{\odot}}\else{M$_{\odot}$\ts}\fi}
\def\cii  {\ifmmode{{\rm [C}{\rm \scriptstyle II}]}\else{[C\ts {\scriptsize II}]\ts}\fi}
\def\ci   {\ifmmode{{\rm [C}{\rm \scriptstyle I}]}\else{[C\ts {\scriptsize I}]\ts}\fi}
\def\m    {\ifmmode{\mu {\rm m}}\else{$\mu$m}\fi}
\def\hi   {\ifmmode{{\rm H}{\rm \scriptstyle I}}\else{H\ts {\scriptsize I}\ts}\fi}
\def\hii  {\ifmmode{{\rm H}{\rm \scriptstyle II}}\else{H\ts {\scriptsize II}\ts}\fi}
\def\nii  {\ifmmode{{\rm [N}{\rm \scriptstyle II}]}\else{[N\ts {\scriptsize II}]\ts}\fi}
\def\oiii {\ifmmode{{\rm [O}{\rm \scriptstyle III}]}\else{[O\ts {\scriptsize III}]\ts}\fi}
\def\hh   {\ifmmode{{\rm H}_2}\else{H$_2$\ts}\fi}
\def\nhh  {\ifmmode{N({\rm H}_2)}\else{$N$(H$_2$)\ts}\fi}
\def\microns {\ifmmode{\mu{\rm m}}\else{$\mu$m\ts}\fi}

\def\lya {\ifmmode{{\rm Ly}{\alpha}}\else{Ly$\alpha$\ts}\fi}
\def\ha   {\ifmmode{{\rm H}{\alpha}}\else{H$\alpha$\ts}\fi}
\def\hb   {\ifmmode{{\rm H}{\beta}}\else{H$\beta$\ts}\fi}

\def\ergcms   {\ifmmode{{\rm erg}\ts{\rm cm ^{-2}}\ts{\rm s ^{-1}}}\else{${\rm erg}\ts{\rm cm ^{-2}}\ts{\rm s ^{-1}}$\ts}\fi}
\begin{document}

   \title{Molecular gas budget of strongly magnified low-mass star-forming galaxies at cosmic noon}

   \author{V. Cat\'an
          \inst{1}
          \and
          J. Gonz\'alez-L\'opez\inst{1,2}
          \and
          M. Solimano\inst{3}
          \and
          L. F. Barrientos\inst{1}
          \and
          A. Afruni\inst{4}
          \and
          M. Aravena\inst{3}
          \and
          M. Bayliss\inst{5}
          \and
          J. A. Hernández \inst{1}
          \and
          C. Ledoux \inst{6}
          \and
          G.  Mahler \inst{7,8,9} 
          \and
          K. Sharon  \inst{10} 
          \and
          N. Tejos \inst{11}}

   \institute{Instituto de Astrof\'isica, Facultad de F\'isica, Pontificia Universidad Cat\'olica de Chile, Av. Vicuña Mackenna 4860, 782-0436 Macul, Santiago, Chile\\
   \email{victoria.catan@uc.cl}
           \and
             Las Campanas Observatory, Carnegie Institution of Washington, 
  Ra\'ul Bitr\'an 1200, La Serena, Chile
           \and
             N\'ucleo de Astronom\'ia de la Facultad de Ingenier\'ia y Ciencias, Universidad Diego Portales, Av. Ej\'ercito Libertador 441, Santiago, Chile
            \and
            Kapteyn Astronomical Institute, University of Groningen, Landleven 12, 9747 AD Groningen, The Netherlands.
            \and
            Department of Physics, University of Cincinnati, Cincinnati, OH 45221, USA
            \and
             European Southern Observatory, Alonso de Córdova 3107, Vitacura, Casilla 19001, Santiago, Chile
             \and
             STAR Institute, Quartier Agora - All\'ee du six Ao\^ut, 19c B-4000 Li\`ege, Belgium
             \and
            Centre for Extragalactic Astronomy, Durham University, South Road, Durham DH1 3LE, UK
            \and
            Institute for Computational Cosmology, Durham University, South Road, Durham DH1 3LE, UK
            \and
            Department of Astronomy, University of Michigan, 1085 South University Avenue, Ann Arbor, MI 48109, USA
            \and
             Instituto de Física, Pontificia Universidad Católica de Valparaíso, Casilla 4059, Valparaíso, Chile}

   \date{Received August 14, 2024; accepted November 4, 2024}

 
  \abstract
   {}
   {The aim of this study is to investigate the molecular gas content of strongly magnified low-mass star-forming galaxies (SFGs) around the cosmic noon period ($z\sim2$) through observations of carbon monoxide (CO) emission lines and dust continuum emission, both of which serve as tracers of molecular gas (H$_2$).}
   {We observed 12 strongly lensed arcs with the Atacama Compact Array (ACA) to detect CO mid-J rotational transitions and dust continuum. Thanks to the strong lensing, we were able to probe the previously understudied low-mass regime. With a compiled set of observations, we recalibrated empirical relations between star formation rate density ($\Sigma_{\rm{SFR}}$)  and the CO line ratios. We derived galaxy properties using spectral energy distribution fitting (SED). We also performed galaxy stacking to combine faint signals. In all cases, molecular gas masses were estimated using both tracers.}
   {We detected CO emission in 3 of the 12 arcs and dust continuum emission in another 3. The obtained H$_2$ masses indicate that most of these galaxies ($M_* < 10^{10.7}$ $\rm{M}_\odot$) have lower molecular gas fractions and shorter depletion times compared to expectations from established scaling relations at these redshifts. We explored several possible explanations for this gas deficit, including uncertainties in mass estimates, effects of low-metallicity environments, larger atomic gas reservoirs in low-mass systems, and the possibility that these represent low-mass analogs of main sequence starburst (MS SBs) galaxies that are undergoing sustained star formation due to gas compaction despite low overall gas fractions.}
   {We conclude that these mass and metallicity regimes present a molecular gas deficit. Our results suggest that this deficit is likely due to a significant amount of atomic gas, which our stacking indicates is about 91\% of the total gas. However, this estimation might be an upper limit, as the possibility remains that our galaxies contain CO-dark gas.}

   \keywords{Star forming galaxies --
                gravitational lensing--
                galaxy evolution -- molecular gas -- CO emission
               }

   \maketitle
%
\section{Introduction}

    Star formation (SF) represents a fundamental astrophysical process mainly occurring within dense, cold molecular clouds. Examining this gas at high redshifts provides insights into galaxy assembly and evolution as it is the primary fuel for SF \citep[e.g.,][]{carilliCoolGasHigh2013,tacconiEvolutionStarFormingInterstellar2020}. The cosmic noon period, which is characterized by a pronounced peak in the cosmic SF rate density around $z\sim$2 \citep[e.g.,][]{madauCosmicStarFormation2014}, is a key epoch for studying the molecular gas (H$_2$) content of galaxies. However, directly studying H$_2$ poses a significant challenge at this time, primarily due to the complexities associated with its detection. Consequently, being able to study H$_2$ during this period holds promise for advancing our understanding of galaxy evolution.
    
    Directly observing H$_2$, the most abundant molecule at high redshift, poses a significant challenge due to its high excitation requirements, even for the lowest transitions \citep[e.g.,][]{carilliCoolGasHigh2013,saintongeColdInterstellarMedium2022}. As a result, astronomers have historically relied on proxies for its study, such as carbon monoxide (CO) \citep[e.g.,][]{carilliCoolGasHigh2013,tacconiEvolutionStarFormingInterstellar2020,saintongeColdInterstellarMedium2022}. There is a preference for this proxy given its abundance and relatively low excitation requirements, facilitated by collisions with $\rm{H}_2$ molecules \citep[e.g.,][]{carilliCoolGasHigh2013}.

    The use of CO as a tracer requires a conversion factor, $\alpha_{\rm{CO}}$ \citep[e.g.,][]{carilliCoolGasHigh2013,tacconiEvolutionStarFormingInterstellar2020,saintongeColdInterstellarMedium2022}, which is defined as 
    
    \begin{equation}\label{eq_alpha}
    M_{\rm{mol}} =  \alpha_{\rm{CO}} {L'}_{\rm{CO(1-0).}}
    \end{equation}
    
    This factor has been observed to change with metallicity, although the exact relationship remains undetermined, and, consequently, several models have been developed to estimate it. Some of these models \citep{wilsonMetallicityDependenceCOtoH1995,magdisGOODSHerschelGastodustMass2011,genzelMetallicityDependenceCO2012,schrubaLowCOLuminosities2012,genzelCombinedCODust2015,accursoDerivingMultivariateACO2017,tacconiPHIBSSUnifiedScaling2018,maddenTracingTotalMolecular2020} are compared in Fig.~\ref{fig_alfa}. Most models show general agreement near solar metallicities but exhibit discrepancies at higher or lower metallicity values. They are also all inversely proportional, as CO gets photo-dissociated at low metallicities \citep{tacconiEvolutionStarFormingInterstellar2020}.

    From the study of CO observations, several authors have established redshift-dependent scaling relations between the molecular gas content and galaxy properties \citep{dessauges-zavadskyMolecularGasContent2015,tacconiPHIBSSUnifiedScaling2018,aravenaALMASpectroscopicSurvey2019,tacconiEvolutionStarFormingInterstellar2020}. Such studies indicate a decreased molecular gas depletion timescale and increased molecular gas fraction with increasing redshift. It is important to note that at high redshift, most studies focus on high-mass galaxies, leaving a gap in our understanding of their low-mass counterparts. As a result, there is a significant lack of research on the molecular gas budget in high-redshift low-mass galaxies, and dedicated research is called for to address this gap.
    
     Gravitational lensing has emerged as an invaluable tool for addressing this deficit, as it facilitates the detection of low-mass galaxies at high redshifts. Strong gravitational lensing by a group or cluster of galaxies can act as a “cosmic telescope,”  magnifying the light from faint background galaxies. The magnification effect makes sources appear more extensive in the sky while preserving their surface brightness, thus allowing us to observe faint galaxies at high redshift, which usually appear as gravitational arcs \citep[e.g.,][]{sharonStrongLensModels2020,sharonCosmicTelescopeThat2022}. Previous works such as \citet{dessauges-zavadskyMolecularGasContent2015} and \citet{saintongeValidationEquilibriumModel2013} studied high-redshift lensed galaxies and obtained their molecular gas contents. These studies used CO as a tracer and focused on star-forming galaxies (SFGs) and found lower depletion times than, and similar gas fractions to, $z\sim$1 main sequence (MS) galaxies. Similarly, \citet{solimanoMolecularGasBudget2021} analyzed four strongly lensed, low-mass galaxies at $z\sim$2 and found relatively low depletion times and gas fractions, which indicated a possible molecular gas deficit in low-mass galaxies.  
     
    Building upon the groundwork laid out by \citet{solimanoMolecularGasBudget2021}, the aim of the present study is to deepen our understanding of the apparent gas deficit found. Our approach involves measuring the CO emission lines of 12 gravitationally lensed arcs with literature-reported stellar masses below $10^{10.5} \rm{M}_{\odot}$ as  \citet{solimanoMolecularGasBudget2021} found they may have different gas reservoirs compared to their high-mass counterparts. The redshifts analyzed range from 1.91 to 3.625, placing them within the cosmic noon period. All galaxies belong to the Sloan Digital Sky Survey Giant Arcs Survey \citep[SGAS;][]{hennawiNewSurveyGiant2008}, making them strongly magnified sources. To enhance the reliability of our findings, we implemented new methodologies for estimating the luminosity conversion ratios and used the method presented in \cite{tsukuiEstimatingStatisticalUncertainty2023} to estimate interferometric errors. We also explored how stacking these galaxies provides information about the average galaxy at these redshifts.  
    
    The paper is organized as follows: In Sect.~\hyperlink{section.2}{2}, we present the sample, the Atacama Compact Array (ACA) observations, and the ancillary data. In Sect.~\ref{analysis}, we present our methods and main results. In Sect.~\ref{discussion}, we discuss and interpret the results. Finally, we report our main conclusions in Sect.~\ref{conclusions}. 
    
    For this study, we employed a flat $\Lambda$CDM cosmology, adhering to the following parameter values: energy density parameter ($\Omega_{\Lambda}$) = 0.7, and matter density parameter ($\Omega_{\rm{M}}$) = 0.3. Additionally, we adopted the initial mass function (IMF) derived by \cite{chabrierGalacticStellarSubstellar2003}.  
    
\section{Observations}
 \begin{figure*}
   \resizebox{\hsize}{!}
            {\includegraphics[width=0.9\hsize]{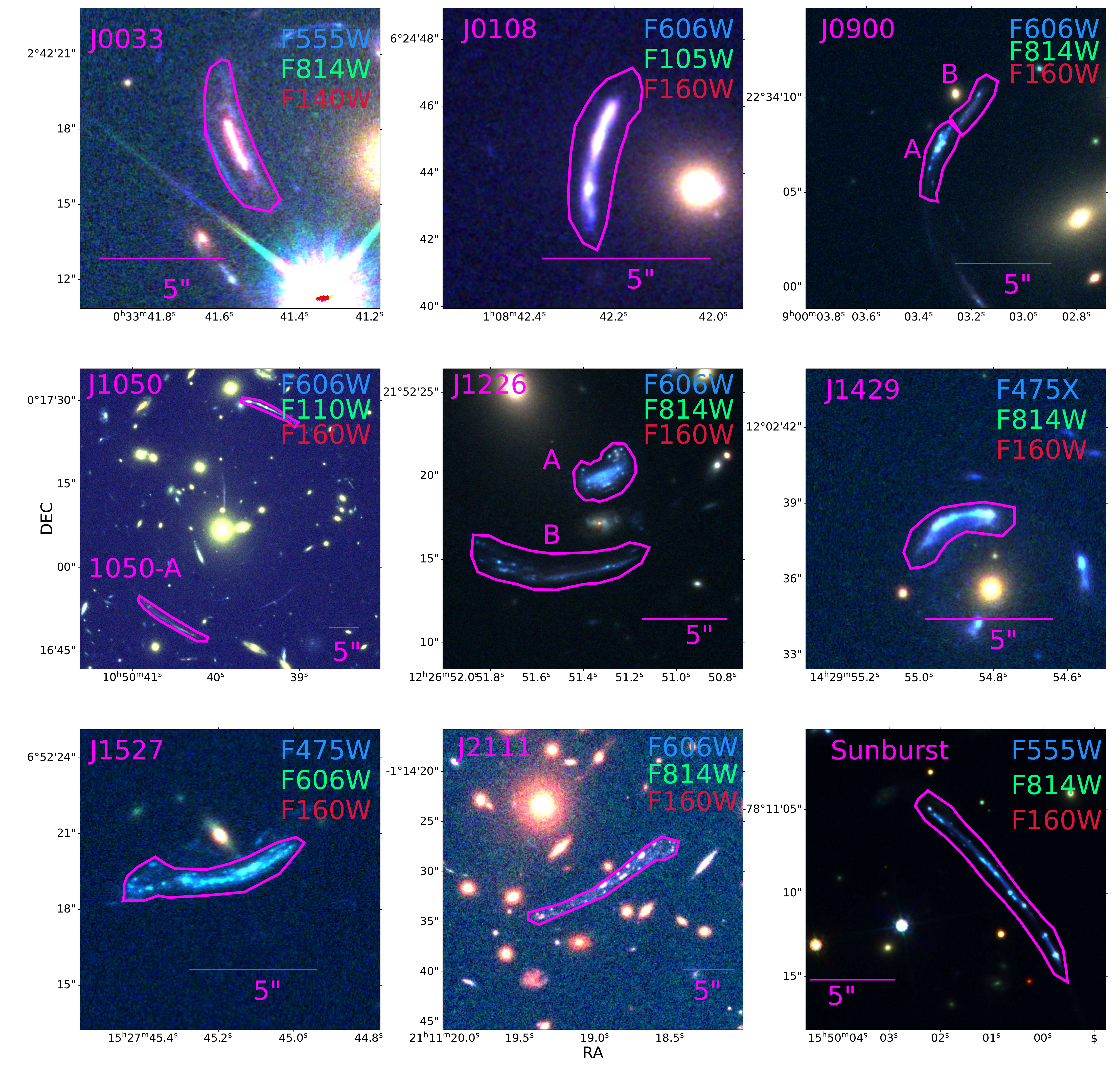}}
      \caption{Color images of gravitational arcs from the HST, labeled with target coordinates and the filters used to create them (F475X (WFC3 UVIS), F555W (WFC3 UVIS), F606W (WFC3 UVIS), F814W(WFC3 IR), F105W(WFC3 IR), F140W(WFC3 IR), F160W(WFC3 IR)). The pink contour indicates the region of the arcs considered to perform HST photometry as indicated in Sect.~\ref{SED_text}. Meanwhile, the pink line indicates the  scale of 5 arcsec.}
         \label{fig_arcos}
   \end{figure*}
   
\subsection{Sample selection}
The sample analyzed in this paper comprises 12 SGAS arcs \citep{hennawiNewSurveyGiant2008}: 4 of these were previously studied by \citet{solimanoMolecularGasBudget2021}, and we selected a further 8 to expand the study. The primary selection criterion was stellar mass, as we aim to investigate the molecular gas content of low-mass galaxies. Consequently, all galaxies in the sample have literature-reported stellar masses of below $10^{10.5} \rm{M}_{\odot}$. Table ~\ref{table_metallicities} reports more details about the galaxies. For the galaxies' metallicities, we used the STARBURST99 (SB99) \citep{leithererStarburst99SynthesisModels1999} values obtained from \citet{chisholmConstrainingMetallicitiesAges2019} and \citet{baylissPhysicalConditionsMetallicity2014}, following the procedure outlined by \citet{solimanoMolecularGasBudget2021}. We assumed a mass-metallicity relation from \citet{sandersMOSDEFSurveyEvolution2021} to estimate sources without previous metallicity measurements, as their masses and redshifts apply to our sample. Though, in some cases, more accurate metallicity values were available, we decided to use SB99 metallicities for all cases to ensure consistency. Overall, we can see that the galaxies are characterized by their low metallicities. 

Figure~\ref{fig_arcos} shows false-color images produced with data obtained with the Hubble Space Telescope (HST). More details about this are presented in Sect.~\ref{sec.HST}. The images show the large angular extension produced by the strong-lensing effect of magnification. The pink lines indicate the scale of 5 arcsec for each image, and the pink contours indicate the regions used for HST photometry as indicated in Sect.~\ref{SED_text}.

\begin{table*}
\caption{Summary of the properties of the galaxies included in this study. }           
\label{table_metallicities}      
\centering          
\begin{tabular}{l l c c c c c}     
\hline\hline       
Arc& Abbreviation  & Coordinates & $z_{\rm{spec}}$ & 12 + Log[O/H]& $\mu$& Discovery\\
\hline                    
 SGASJ003341.5+024217     & J0033          & 00:33:41.55            & 2.3900$^{(1)}$   & 8.61$\pm$0.02$^{(2)}$ & 24$^{+9}_{-5}$  & (1)    \\
                          &                & +02:42:16.58$^{(1)}$  &                &               &              &\\
SGASJ010842.2+062444      & J0108          & 01:08:42.21            & 1.9099$^{(1)}$   & 8.39$\pm$0.03$^{(2)}$ & 5$^{+4}_{-2}$           & (3)\\
                          &                & +06:24:44.41$^{(1)}$   &                &               &              & \\
SGASJ090003.3+223408      & J0900A/J0900B  & 09:00:03.33            & 2.0323$^{(1)}$   & 8.27$\pm$0.02$^{(2)}$ &6$^{+4}_{-2}$/8$^{+5}_{-3}$           & (7) \\
                          &                & +22:34:07.57$^{(1)}$   &                &               &              &  \\
SGAS J105039.6+001730 C.1 & J1050          & 10:50:39.398           & 3.6258$^{(10)}$ & 8.30$\pm$0.09$^{(10)}$& 72$^{+9}_{-10}$&  (10)  \\
                          &                &  +00:17:28.78$^{(6)}$ &                &               &              &\\
SGAS J105039.6+001730 A.1 & J1050A         & 10:50:40.358           &2.4034$^{(10)}$&   8.51$\pm$0.02    &19$\pm$1           & (10)\\
                          &                &  +00:16:48.54$^{(6)}$ &                &               &              &  \\
SGASJ122651.3+215220      & J1226A/J1226B  & 12:26:51.32            & 2.9252$^{(1)}$   & 8.22$\pm$0.03$^{(2)}$  &84$^{+13}_{-10}$/31$^{+6}_{-4}$     & (8) \\
                          &                & +21:52:20.00$^{(1)}$   &                &               &              &\\
BG J1429+1202 B           & J1429          &  14:29:54.857$^{(1)}$ &  2.8241$^{(1)}$&  8.45$\pm$0.03$^{(2)}$& 6$^{+2}_{-1}$        & (12)\\
                          &                &  +12:02:38.68          &                &               &              & \\
SGAS J152745.1+065219     & J1527          & 15:27:45.17            &  2.7627$^{(1)}$& 8.38$\pm$0.04$^{(2)}$ & 12$^{+2}_{-1}$       &   (11) \\
                          &                & +06:52:19.17$^{(11)}$ &                &               &              & \\
SGASJ211118.9$-$011431    & J2111          & 21:11:18.95            & 2.8590$^{(1)}$   &   8.51$\pm$0.02     & 27$^{+10}_{-10}$          & (9)\\
                          &                & $-$01:14:31.44$^1$     &                &               &              & \\
PSZ1G311.6518.4           & Sunburst       & 15:49:59.83            &  2.3709$^{(5)}$  &   8.43$\pm$0.03$^{(2)}$& 171$^{(13)}$& (4)\\
                          &                & $-$78:11:13.2$^{(4)}$  &                &               &              & \\
\hline                  
\end{tabular}
\tablefoot{The first column indicates the complete name of the arc, while the second column shows the abbreviation used in this work. Column 3 indicates the coordinates of each arc, and column 4 the spectroscopic redshifts. The fifth column presents the metallicity of the galaxies, and column 6 shows the magnification of each of the studied arcs. Column 7 indicates the discovery paper of the galaxy.}
\tablebib{$^{(1)}$~\citet{rigbyMagellanEvolutionGalaxies2018},$^{(2)}$~\citet{chisholmConstrainingMetallicitiesAges2019},$^{(3)}$~\citet{rigbyLackCorrelationMg2014},$^{(4)}$~\citet{dahleDiscoveryExceptionallyBright2016},$^{(5)}$~\citet{rivera-thorsenSunburstArcDirect2017},$^{(6)}$~\citet{sharonStrongLensModels2020},$^{(7)}$~\citet{diehlSloanBrightArcs2009},$^{(8)}$~\citet{koesterTWOLENSEDLYMAN2010},$^{(9)}$~\citet{oguriSubaruWeakLensing2009},$^{(10)}$~\citet{baylissPhysicalConditionsMetallicity2014},$^{(11)}$~\citet{koesterTWOLENSEDLYMAN2010},$^{(12)}$~\citet{marques-chavesDiscoveryVeryBright2017},$^{(13)}$~\citet{solimanoMolecularGasBudget2021}}
\end{table*}

\subsection{ACA data}

This study comprises observations from three projects conducted with the ACA: 2018.1.01142.S (PI: Gonz\'alez-L\'opez), 2021.2.00092.S (PI: Solimano), and 2022.1.00916.S (PI: Solimano). This array consists of 12 radio antennae, each of 7 m in diameter, operating in interferometric mode. This generates images with a larger synthesized beam than the 12 m array but with superior sensitivity to large-scale structures, which makes it an effective tool for detecting the extended emission of molecular lines and dust continuum in giant arcs.
In our study, each gravitationally lensed arc was observed in two separate bands: one for detecting CO line emission and the other for obtaining the dust continuum. We targeted a single mid-J CO line for every source, with J=3, 4, or 5 depending on redshift. The observations focused on bands 3 and 4 for CO, depending on the galaxy's redshift, and bands 6 and 7 for the dust continuum. The bands used to study the continuum were not chosen to include the expected emission lines. However, emission lines were searched for in the cubes, and no significant emissions were detected. Further details of the observations are provided in Table~\ref{table_alma}, where the central frequency is the midpoint between the lowest and highest frequency observed in each visibility.

\begin{table*}
\caption{ACA observations information.}
\label{table_alma}      
\centering                          
\begin{tabular}{l c c c c c c }        
\hline\hline                 
Arc& \multicolumn{2}{c}{Central frequency} & \multicolumn{2}{c}{rms} & \multicolumn{2}{c}{Synthesized beam}\\ 
& CO band & Dust band& CO band & Dust band& CO band & Dust band\\
& [GHz] & [GHz]& [mJy/beam] & [mJy/beam]& [arcsec x arcsec] & [arcsec x arcsec]\\
\hline                        
J0033 & 142 & 344 &1.7& 0.2 &13.3$\times$7.2& 4.7$\times$3.3\\      
J0108 & 153 & 344&1.6&0.3&10.7$\times$6.6& 4.2$\times$3.0 \\
J0900& 146 & 344 &1.8& 0.2&10.7$\times$8.4& 4.6$\times$4.0    \\
J1050/J1050A& 95 & 231 &1.3&0.1&16.8$\times$10.3& 7.2$\times$4.6    \\
J1226& 140 & 344 &1.9&0.1&10.0$\times$9.0&   4.8$\times$3.8  \\
J1429& 96 &   310 &2.2&0.6&18.5$\times$12.2& 5.7$\times$3.5   \\
J1527& 98 & 308  &1.5&0.2&18.5$\times$11.7&  6.1$\times$3.2  \\
J2111& 96 &  344 &1.0&0.1&10.5$\times$8.4&  5.6$\times$4.2  \\
Sunburst & 141 & 344 &1.5&0.2&11.5$\times$10.9&   5.0$\times$4.7 \\ 
\hline                                   
\end{tabular}
\tablefoot{Column 1 indicates the central frequency of the observations for both the CO band (bands 3 or 4) and the dust band (bands 6 or 7). Column 2 shows the rms of the observations; for the CO bands, this was estimated for one channel of 30 kms$^{-1}$ in which emission was expected, and for the dust band, it was calculated using the MFS images, which covered 7.5 GHz. Column 3 indicates the sizes of the synthesized beams for each data product.}
\end{table*}
The ACA observations were calibrated using Common Astronomy Software Applications package \citep[{\sc CASA},][]{casateamCASACommonAstronomy2022} versions 6.2.1.7 and 6.4.1.12 for the different projects. The calibration was done using the pipeline corresponding to each Atacama Large Millimeter/submillimeter Array (ALMA) cycle. No extra flagging was used in addition to the ones identified by the observatory and the automatic flagging identified by the pipeline. Inspection of the pipeline's weblog and products showed no features or anomalies that could point to problems with the calibration.  

The data were processed into spectral cubes and multi-frequency synthesis (MFS)
images using {\sc CASA}. Spectral cubes were made using 60 channels of 30 km s$^{-1}$ using the spectral windows (spws) close to the expected line. We applied a Hogbom deconvolution algorithm and a natural weighting. In contrast, the MFS images combine all channels into a single continuum image covering a bandwidth of 7.5 GHz for bands 6 and 7, yielding a product suited for searching for and measuring dust thermal emissions. We cleaned the image or cube using a three-sigma threshold when emission was found; for this, we also interactively created manual masks, providing a prior of the source position. We created cubes only for bands 3 and 4, while MFS images were created for all bands. However, it is important to mention that for bands 3 and 4, the channels where emission was expected were not considered for the MFS. 

\subsection{Ancillary data}
\subsubsection{HST data} \label{sec.HST}

We analyzed HST images obtained through various filters and instruments to construct spectral energy distributions (SEDs) for each galaxy. The data were retrieved via the Mikulski Archive for Space Telescopes (MAST). Different instrument configurations were employed, depending on the galaxy, with the Wide Field Camera 3 (WFC3) used for infrared (IR) and ultraviolet-visible (UVIS) filters, the Advanced Camera for Surveys (ACS) for optical filters, and the Wide Field Planetary Camera 2 (WFPC2) for additional filters (F450W, F606W, F814W). The High-Level Science Products (HLSPs) released by \citet{sharonStrongLensModels2020} were used for most arcs in this analysis. In cases where the HLSP data were unavailable, data reduced by the Hubble Advanced Products (HAP) pipeline were used instead. For the Sunburst Arc, the individual HAP exposures were combined using the AstroDrizzle package, while for arc J1226, the HLSP data was obtained from the TEMPLATES program Sharon et al. (in prep.).

\subsubsection{James Webb Space Telescope  data}

We complemented the above data with James Webb Space Telescope (JWST) observations for arc J1226 belonging to the TEMPLATES program \citep{RigbyTEMPLATESTargetingExtremely2017}. The NIRCAM observations used filters F115W, F150W, F200W, F277W, F356W, and F444W, while MIRI observations employed filters F560W and F770W. These images were reduced using scripts provided by \citet{spilkerSpatialVariationsAromatic2023}. The data used for the SED fitting are in Zenodo (See Sect.~\ref{data}).

\subsubsection{Lens models}
We estimated magnification factors using parametric lens models constructed with the \textsc{Lenstool} software \citep{julloMultiscaleClusterLens2009}. We used the models published by \citet{sharonStrongLensModels2020} for most of the fields in our sample. For J0033, J1226, and the Sunburst arc, we used the models published by \cite{fischerSpatiallyResolvedOutflows2019}, Sharon et al. (in prep.), and \cite{sharonCosmicTelescopeThat2022}, respectively. Meanwhile, the only published model for J1429 was derived from seeing-limited imaging \citep{marques-chavesDiscoveryVeryBright2017a}, and therefore suffers from significant uncertainties. Here, we developed a new lens model for J1429 based on the available HST data.

As the lensing of the source in the J1429 field is dominated by a single galaxy (the brightest cluster galaxy, BCG), we chose to model the system as a single pseudo-isothermal elliptical mass distribution (PIEMD) at $z_\mathrm{lens}=0.55$ plus an external shear, instead of adding the individual cluster members as separate potentials. We keep the central coordinates of the potential fixed at $\alpha=217.478368^\circ$, $\delta=12.043233^\circ$, corresponding to the center of the BCG. The model is constrained by three quadruply imaged regions of the source we identified in the color HST image (Fig.~\ref{fig_arcos}). We then optimized the seven free parameters of the model using \textsc{Lenstool}'s  Markov Chain Monte Carlo (MCMC) sampling. The marginal percentiles from the posterior distribution samples are presented in Table~\ref{tab:sgasj1429_lensmodel}.

\begin{table}[!htb]
    \centering
    \caption{Lens model parameters for the J1429 field.}
    \label{tab:sgasj1429_lensmodel}
    \begin{tabular}{lc}
         \hline
         \hline
         Parameter &   16-50-84 PDF percentiles \\
         \hline
         \multicolumn{2}{c}{PIEMD potential} \\
         $\hat{\epsilon}$ & [0.19, 0.26, 0.96] \\
         $\theta_{\hat{\epsilon}}$ ($^\circ$) &  [57.06, 68.63, 71.55] \\
         $r_{\mathrm{core}}$ (arcsec) & [0.19, 0.31, 0.68] \\
         $\sigma_\mathrm{dPIE}$ (km/s) &  [336.4, 344.2, 378.9] \\
         $r_{\mathrm{cut}}$ (arcsec) & [19.0, 20.0, 47.65] \\
         \hline
         \multicolumn{2}{c}{External shear} \\
         $\gamma$ & [0.171, 0.217, 0.306] \\
         $\theta_\mathrm{shear}$ ($^\circ$) &  [-10.5, -7.9, -4.4] \\
         \hline
    \end{tabular}
\end{table}

We derived deflection matrices for all the arcs in our sample at each of the 100 model realizations drawn from the MCMC. We used the angular deflections prescribed by these matrices to bring our photometric apertures to the source plane. We then computed the mean magnification factor inside the aperture as the ratio between its image-plane and source-plane areas, $\mu$. Finally, we used the median of the 100 samples as the nominal value and we adopted $\sigma_{\mu}\equiv(p_{84} - p_{16})/2$ as the uncertainty, where $p_{84}$ and $p_{16}$ are the 84\textsuperscript{th} and 16\textsuperscript{th} percentiles, respectively. The magnifications obtained in this way are reported in Table~\ref{table_metallicities} and applied to de-lensed flux-dependent quantities throughout the paper.

\section{Data analysis and results} \label{analysis}

\subsection{Emission lines}
\begin{figure}
\centering
\includegraphics[width=0.9\hsize]{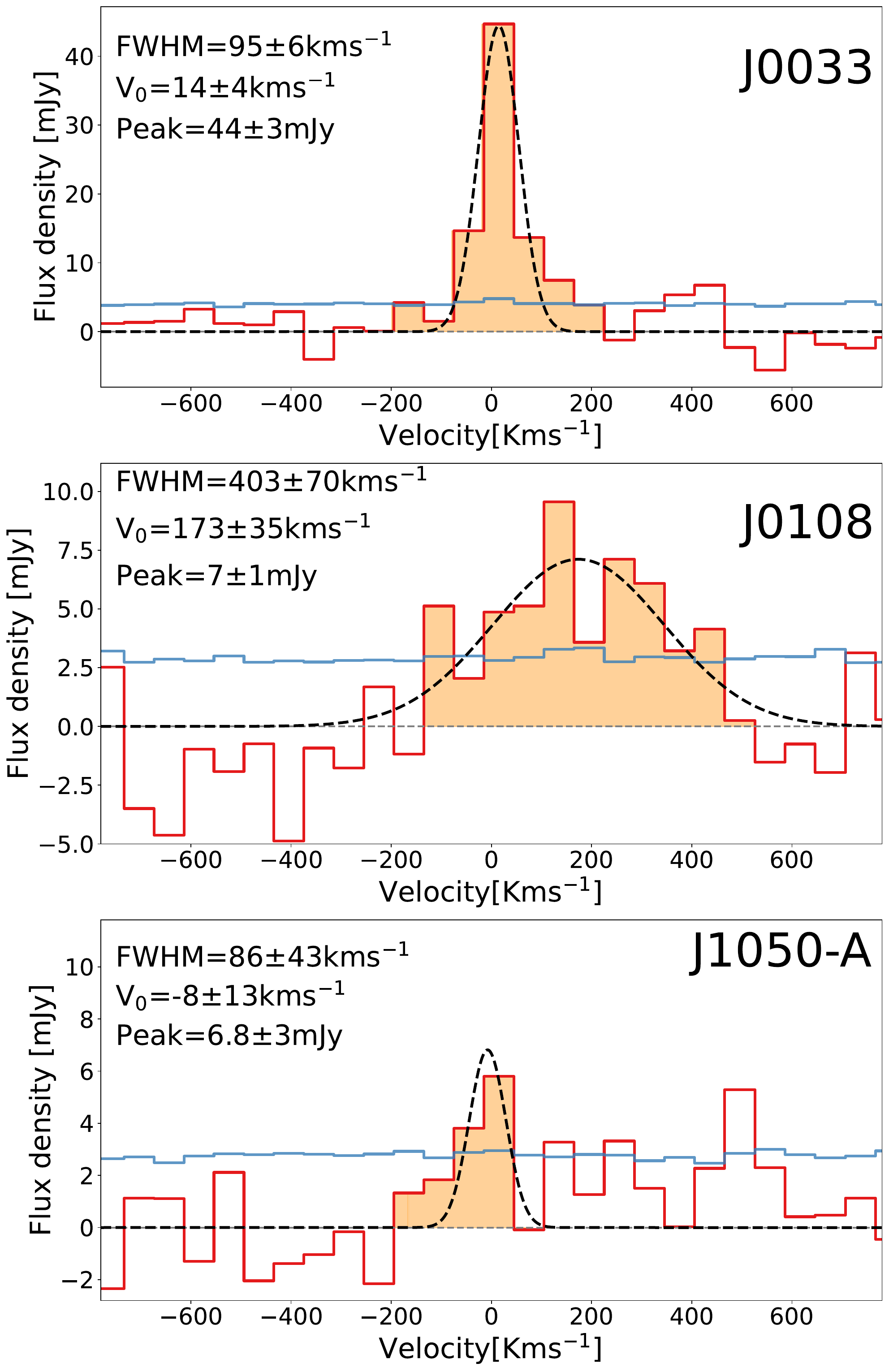}
\caption{Spectra of the CO emission detections. The galaxies shown are J0033 (top), J0108 (middle), and J1050A (bottom). All plots are centered to an optical velocity of zero. The orange channels correspond to the channels collapsed to create the zeroth moment. A Gaussian fit (black curve) is overlaid to determine the fit parameters, the full width half maximum (FWHM) in km$\rm{s}^{-1}$, the maximum amplitude in mJy, and the velocity in which the Gaussian is centered. The blue lines represent the 1$\sigma$ noise level.}
\label{fig_spectras}
\end{figure}

We conducted a search for CO emission lines in all the targets at the expected redshifted frequencies, manually examining the cubes with a resolution of 30 km$\rm{s}^{-1}$. We found three possible lines corresponding to arcs: J0033, J0108, and J1050A. In Fig.~\ref{fig_spectras}, the spectra of the extended emission of these arcs are depicted. In the figure, the blue line represents the 1$\sigma$ noise level. We considered the emissions to be significant if the integrated flux exceeded the 2$\sigma$ level. \\

The zeroth moment maps for these arcs were obtained by collapsing the orange channels. A velocity range of 210 km$\rm{s}^{-1}$ centered on the systemic velocity was collapsed for arcs without detections. This velocity range was chosen due to its similarity to the collapsed range for detecting J0033. The resulting zeroth moment maps were individually analyzed to derive the integrated CO fluxes of the galaxies. Two different procedures were employed, depending on whether the source was resolved or unresolved. For unresolved sources, the emission was determined as the peak value, with the images' root mean square (rms) serving as its uncertainty. For resolved sources, aperture photometry was performed within the galaxy region covering 1$\sigma$ emission, and uncertainties were calculated using the ESSENCE code presented in \citet{tsukuiEstimatingStatisticalUncertainty2023}. This code estimates uncertainties using the autocorrelation function between pixels. The derived integrated fluxes were then used to calculate the CO luminosities of each galaxy, as presented in Table ~\ref{table_emission}.\\

\begin{table}
\caption{Integrated CO line fluxes and derived CO luminosities for each galaxy.}             
\label{table_emission}      
\centering                          
\begin{tabular}{l c c c} 
\hline\hline 
Arc           &CO    & $\mu\Delta \nu S_{\nu}$ & $\mu {L'}_{\rm{CO}}$\\
&Transition&[Jy $\cdot$ Km $\rm{s}^{-1}$]&[$10^{10}$K km $\rm{s}^{-1}$ $\rm{pc}^{2}$]\\
\hline                

J0033  & (4-3)  & 2.0$\pm$0.1   & 3.4$\pm$0.2 \\
J0108  & (4-3)  &  1.3$\pm$0.4  & 1.4$\pm$0.4 \\
J0900A & (4-3)  & <0.5 & <0.6\\
J0900B & (4-3)  & <0.5  & <0.6\\
J1050  & (4-3) & <0.3 & <1.2\\
J1050A & (3-2) & 0.4$\pm$0.1 & 1.2$\pm$0.3 \\
J1226A & (5-4)  & <0.4 &<0.7  \\
J1226B & (5-4) & <0.2 & <0.4 \\
J1429  & (3-2) & <0.4 & <1.7 \\
J1527  & (3-2) & <0.4 & <1.6 \\
J2111  & (3-2)  & <0.2 & <0.9 \\
Sunburst   &(4-3)& <0.4& <0.6   \\   
\hline                                   
\end{tabular}
\tablefoot{Listed are the galaxy names, CO transition line, unmagnified integrated flux (Jy km$\rm{s}^{-1}$), and unmagnified calculated CO luminosities ($\mu {L}'_{\rm{CO}}$ in units of $10^{10}$ K km $\rm{s}^{-1} \rm{pc}^{2} $). Statistical uncertainties on the measured values are included, with upper limits being considered at the 3$\sigma$ significance level.}
\end{table}

\subsection{Continuum data}
\begin{figure*}
\resizebox{\hsize}{!}
        {\includegraphics[width=0.9\hsize]{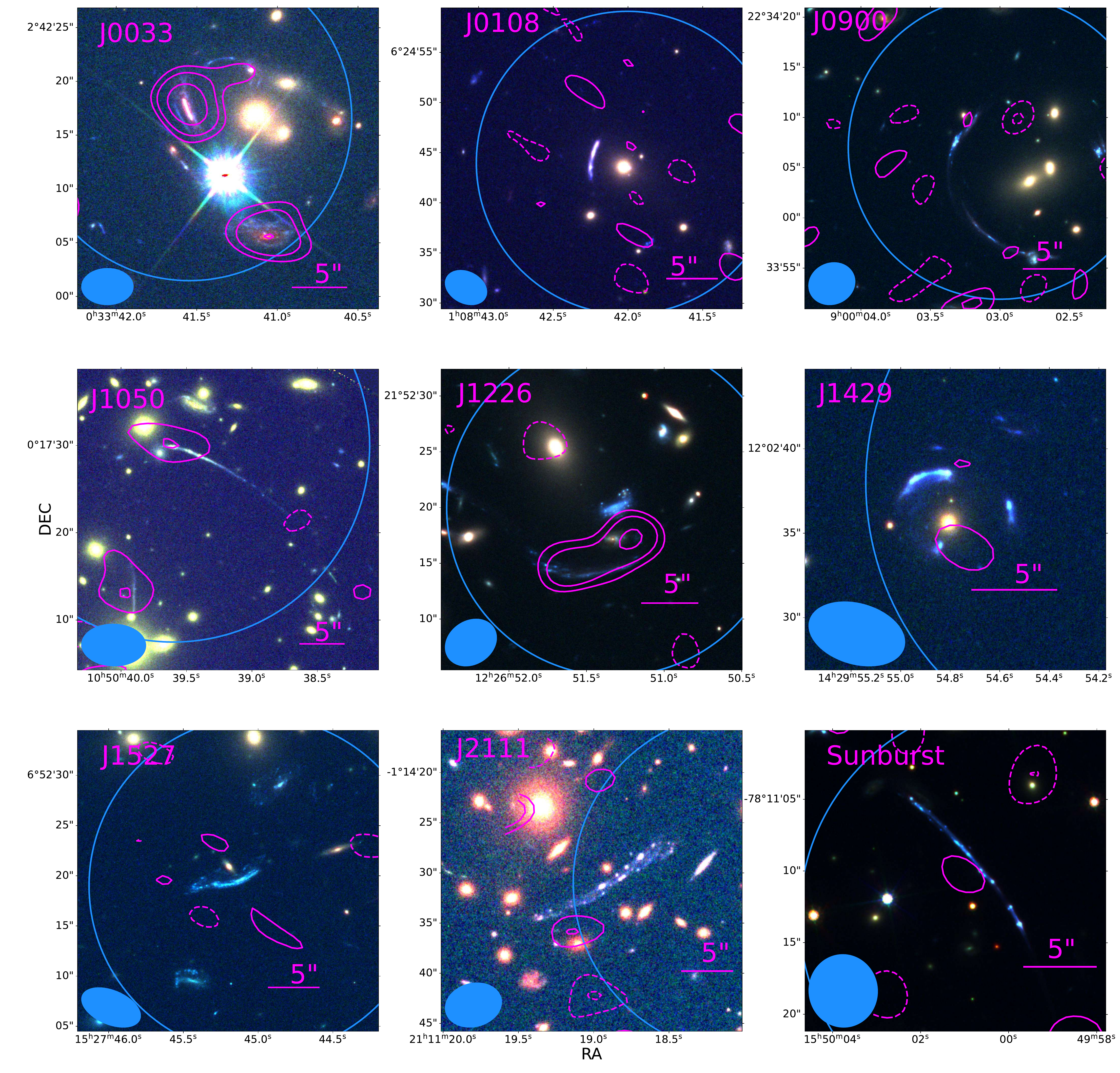}}
  \caption{HST color images of gravitational arc systems with overlaid contours showing dust continuum emission observed with ACA at different $\sigma$ levels (-5, -3, -2, 2, 3, 5). Dashed lines correspond to negative values while solid lines are positive values. The sizes of the synthesized beams are indicated in the bottom left corners of the images. Additionally, the large circles represent the ACA's primary beam at a value of 0.5 for the respective target fields.}
     \label{fig_cont}
\end{figure*}

We investigated the dust continuum emission from the galaxies using MFS images, which considered all spws. This enabled the extraction of integrated continuum fluxes for each arc. Continuum detections were identified when the emission exceeded a 2$\sigma$ significance level. The arcs and their dust continuum are depicted in Fig.~\ref{fig_cont}. Flux measurements for the arcs were conducted using the same procedures employed for obtaining the fluxes in the zeroth moment maps. The derived continuum flux densities, expressed in mJy, are listed in Table ~\ref{table_params}.\\

\subsection{SED fitting} \label{SED_text}
In this study, we employed the Code Investigating GALaxy Emission (CIGALE; \citealt{boquienCIGALEPythonCode2019}) to perform SED fitting. CIGALE operates on the principle of energy balance, where the energy absorbed by dust is re-emitted in the infrared spectrum. We provided CIGALE with each galaxy's redshift and HST, ALMA, and JWST photometry. Specifically, we adopted a delayed SF history and used the \citet{bruzualStellarPopulationSynthesis2003} simple stellar population model. Additionally, we accounted for nebular emission and dust attenuation, employing the \citet{charlotSimpleModelAbsorption2000} attenuation law and the dust templates proposed by \citet{daleTwoparameterModelInfrared2014}. \\

\begin{figure}
   \hspace*{-0.5cm}  
            {\includegraphics[width=0.45\textwidth]{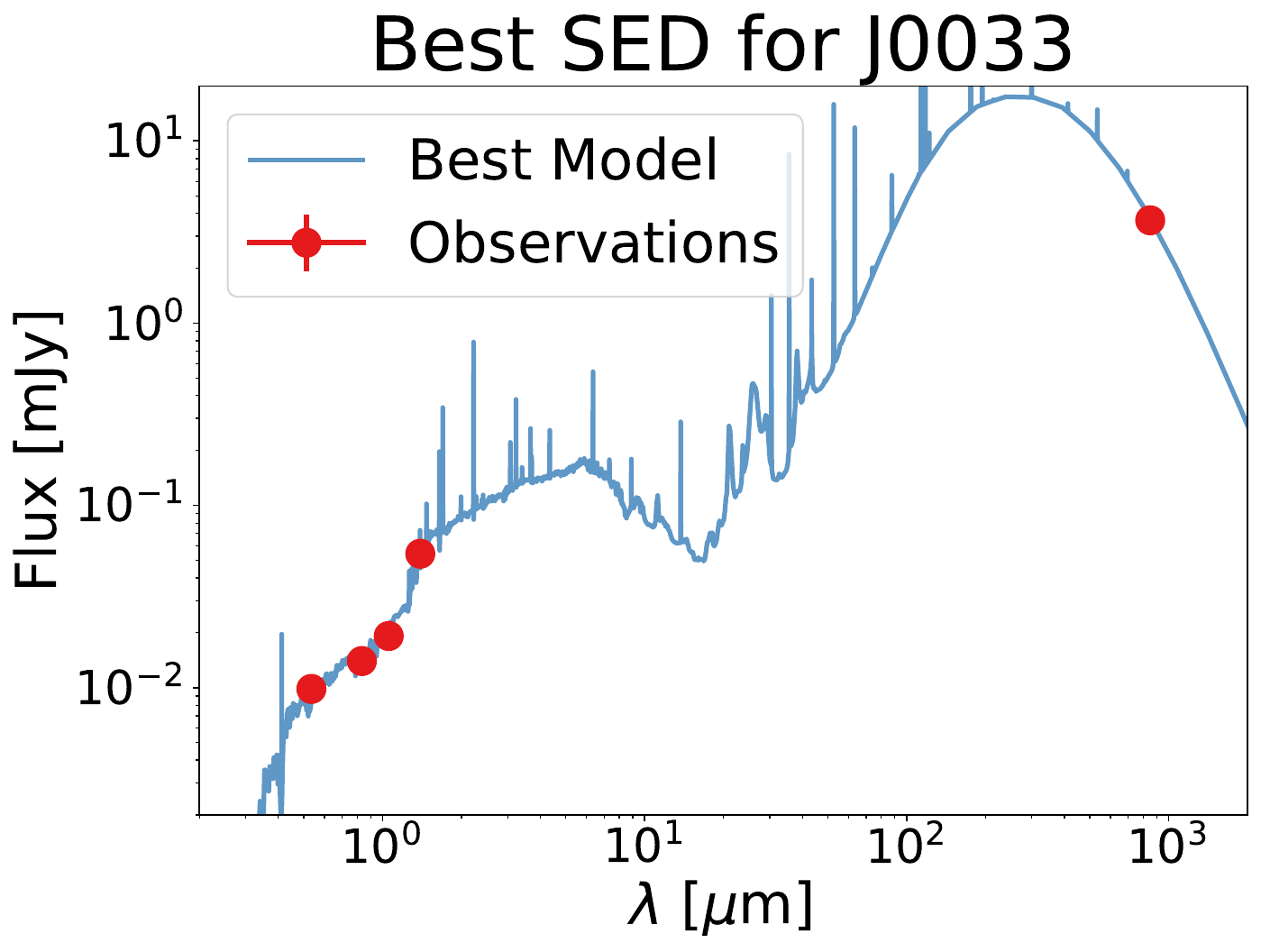}}
      \caption{Best SED fit for galaxy J0033 using HST and ACA photometry obtained with CIGALE.}
         \label{SED_1}
   \end{figure}

Our approach involved two stages of SED fitting for each galaxy: one considering the entire arc and another focusing on resolved properties. The latter enabled a robust estimation of the surface SF rate density ($\Sigma_{\rm{SFR}}$) for each region. Aperture photometry was performed on HST and JWST images at various wavelengths for the whole arc, supplemented by ACA band 6 or 7 when available. These apertures were created to cover the highest percentage of the galaxy's flux in the reddest available HST filter and are presented in Fig.~\ref{fig_arcos}. In Zenodo (See Sect.~\ref{data}), the flux percentages covered in each case are available. An illustrative example of a SED fit is presented in Fig.~\ref{SED_1}, and the details of the  parameters used are available in Zenodo (See Sect.~\ref{data}). Table ~\ref{table_params} presents the values derived from the fit.\\

For the resolved analysis, we calculated the SF rate (SFR) for several circular apertures and normalized them by their corresponding area to derive $\Sigma_{\rm{SFR}}$ values. This approach facilitated the characterization of a $\Sigma_{\rm{SFR}}$ distribution across different regions of the galaxies. We considered the median of the distributions as representative values of the total $\Sigma_{\rm{SFR}}$ of each galaxy.\\

\begin{table}
\caption{Dust continuum fluxes obtained from ACA data and the parameters obtained from the CIGALE SEDs.}             
\label{table_params}      
\centering          

\begin{tabular}{l c D{,}{\pm}{-1} D{,}{\pm}{-1} D{,}{\pm}{4.4} }
\hline\hline 
Arc&$\mu S_{\nu}$
& \multicolumn{1}{c}{{Log(${M}_{*}$)}}
& \multicolumn{1}{c}{{Log(SFR)}}
& \multicolumn{1}{c}{{$\Sigma_{\rm{SFR}}$}}\\
&[mJy]& \multicolumn{1}{c}{[$\rm{M}_{\odot}$]}
& \multicolumn{1}{c}{[$\rm{M}_{\odot}\rm{yr}^{-1}$ ]}
& \multicolumn{1}{c}{[$\rm{M}_{\odot}\rm{yr}^{-1}\rm{kpc}^{-2}$]}\\
\hline                       
J0033 &3.7$\pm$0.4&10.7,0.2&1.8,0.2&1.3,0.3\\ 
J0108&<0.4&10.4,0.3&1.7,0.3&0.23,03.\\ 
J0900A&<0.5&9.7,0.2&1.2,0.2&0.8,0.2\\ 
J0900B&<0.4&9.9,0.2&1.0,0.3&1.2,0.5\\ 
J1050 &0.4$\pm$0.1&9.8,0.2&1.2,0.1&1.5,0.2\\ 
J1050A& - &10.0,0.2&1.5,0.2&0.35,0.03\\ 
J1226A&<0.3&9.5,0.1&0.8,0.1&1.4,0.1\\ 
J1226B&0.8$\pm$0.2&10.0,0.1&1.3,0.2&0.49,0.03\\ 
J1429&<2.5&10.6,0.3&1.9,0.2&46,11\\ 
J1527&<0.5&10.2,0.4&1.5,0.2&10,4\\ 
J2111&<1.0&9.9,0.3&1.2,0.2&0.23,0.01\\ 
Sunburst&<1.0 &9.5,0.1&0.9,0.1&0.52,0.03\\ 
\hline                  
\end{tabular}
\tablefoot{Column 2 presents the magnified dust continuum values. Columns 3, 4, and 5 show demagnified values for stellar mass, SFR, and($\Sigma_{\rm{SFR}}$), respectively.}
\end{table}

\subsection{Dust mass}

We used the dust continuum observations to constrain the dust mass of our targets. Our estimates were made by applying the equations in \citet{caseyPhysicalCharacterizationUnlensed2019}:
\begin{equation}
{M}_{\rm{dust}} = \frac{{S}_{\nu_{\rm{obs}}} {D}_{\rm{L}}^2 (1+{z})^{-(3+\beta)}}{\kappa (\nu_{\rm{ref}}) {B}_{\nu}(\nu_{\rm{ref}},{T}_{\rm{d}})} \left(\frac{\nu_{\rm{ref}}}{\nu_{\rm{obs}}}\right)^{2+\beta} \left(\frac{\Gamma_{\rm{RJ}}(\rm{ref},0)}{\Gamma_{\rm{RJ}}}\right)
    \label{casey2}
,\end{equation}
where $\kappa$=1.3$\rm{cm}^2 \rm{g}^{-1}$ represents the dust-mass absorption coefficient,\ $\beta$=1.8 denotes the emissivity spectral index, and $T_d$=25K signifies the global mass-weighted dust temperature \citep{scovilleISMMassesStar2016}. The $\Gamma_\mathrm{RJ}$ values serve as corrections to the Planck function, $B_\nu$, accounting for the Rayleigh-Jeans deviation \citep{scovilleISMMassesStar2016}. These values can be estimated using Equation ~\ref{scoville}, where $h$ denotes Planck's constant and $k$ represents the Boltzmann constant. In this equation, we also use $T_d$=25K:
\begin{equation}
\Gamma_{\rm{RJ}}({T}_{\rm{d}}, \nu, {z}) = \frac{{h} \nu (1+{z})/{k} {T}_{\rm{d}}}{{e}^{{h} \nu (1+{z})/{k} {T}_{\rm{d}}}-1}
    \label{scoville}
.\end{equation}
In the upper section of Equation ~\ref{casey2}, $\Gamma_\mathrm{RJ}$ is computed using the reference frequency and a redshift of 0, whereas, in the lower section, it is calculated using the observed frequency and redshift. It is important to note that this equation is applicable only when $\lambda_\mathrm{rest} > 250\,\mu$m, as this is the Rayleigh-Jeans tail, where dust is optically thin \citep{scovilleISMMassesStar2016}, limiting our ability to estimate the dust mass in Band 7 for certain arcs. However, the dust temperature is higher in the case of low-metallicity galaxies, which displaces the Rayleigh-Jeans tail towards shorter wavelengths, allowing us to use this method for more galaxies of our sample \citep{saintongeValidationEquilibriumModel2013}. We considered the limit J1226A/B as they have rest wavelengths of 222$\mu$m. The only galaxy with a higher redshift is J1050. However, as it was observed at 1.25mm, it can still be studied in this context.\\

\subsection{Stacking}
  
Given our sample size, we were able to stack the CO and dust continuum emissions for our observations. This allowed us to study the average behavior of our sample. In all cases, we observed that the signal-to-noise ratio was optimized when galaxies were weighted by $1/\sigma^2$, where $\sigma^2$ represents the data variance. As the galaxies were gravitationally lensed, we applied a magnification correction to all fluxes and images before stacking them. This ensures that our final results represent the intrinsical properties of the galaxies for each stack.\\

\begin{figure}
\centering
\includegraphics[width=0.9\hsize]{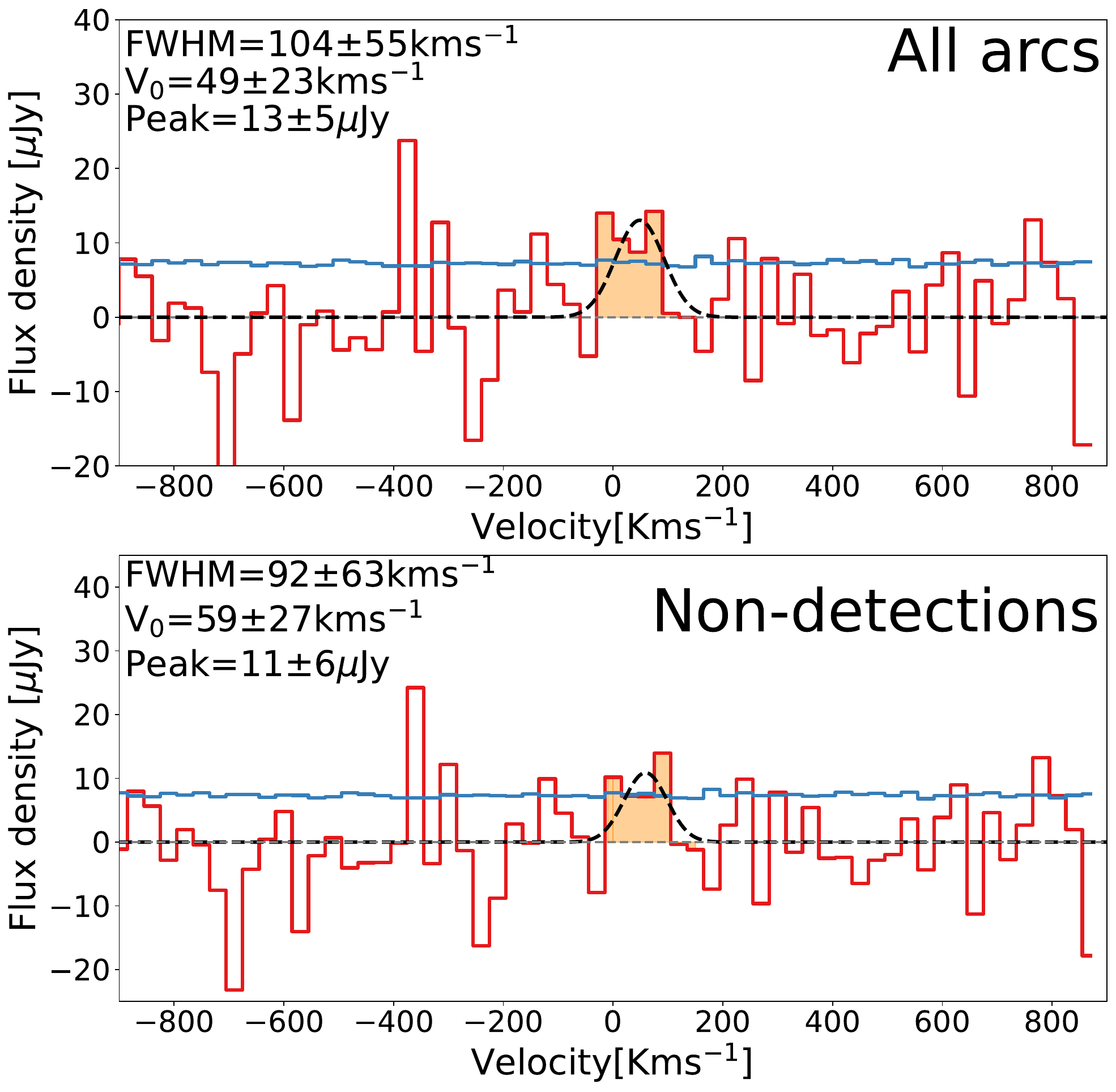}
   \caption{ Stacked CO spectra. The top panel shows the CO stack for all the arcs, while the bottom panel shows the CO stack with no detection. A Gaussian fit was performed in both cases and is overlaid in black. The blue lines represent the 1$\sigma$ noise level, while the orange region indicates the channels collapsed to obtain the integrated fluxes.}
    \label{sepc_stack}
\end{figure}

For CO emissions, we obtained the spectra of a single central pixel for each galaxy, which we then stacked. We also considered the possibility of performing cube stacking; however, this was not done due to the different beam sizes of the observations. Using a single pixel does not consider the extended emission of the arcs; thus, as an alternative to cube stacking, we obtained the spectra of the flux densities  for the extended emission of the arcs. However, we found that the stacking signal became diluted by noise by doing this, and so we decided to use the single-pixel spectra. Primary beam corrections were performed for each spectrum, and the spectral axis was converted to velocity, considering the center of the line as the rest frequency. We first stacked all the arcs and then repeated the process, excluding galaxies J0033, J0108, and J1050A, which presented detections. This provides insight into the behavior of the undetected galaxies. The samples are referred to as All Arcs and Non-detections, respectively, hereafter. Table ~\ref{table_stacking} reports the obtained fluxes for each case. Both stacked spectra can be observed in Fig~\ref{sepc_stack}; here, the two spectra are similar, which indicates that the same spectrum dominates both stackings. We considered emission to be significant if the integrated flux density of the orange area surpassed 1.5$\sigma$, making them tentative detections. We used the individual galaxy properties and the corresponding weights to obtain representative properties for each stacking, which were then used to estimate the molecular masses. For this, we calculated a weighted average of the galaxy properties using $1/\sigma^2$ as the weight. All values derived from the stacks are presented in Table ~\ref{table_stacking}.\\

It is essential to mention that these results could be affected by differences between spectroscopic and CO redshifts. We can see that this effect is present in Fig.~\ref{fig_spectras}, and so it could also be affecting other arcs. If this were the case, we would expect a higher integrated flux. However, as we obtain a low signal-to-noise ratio, even when including significant detections, we do not expect the molecular gas mass to vary significantly in this case.\\

Regarding the continuum, we performed single-pixel stacking. Observations were divided into bands 3 and 4 (B3/4) and 6 and 7 (B6/7); as a final result, we obtained three stacks. First, we stacked all the galaxies for bands 3 and 4, referred to as Continuum B3/4 all arcs. Then we did the same with bands 6 and 7, referred to as Continuum B6/7 all arcs. Finally, we removed all arcs with detections in bands 6 and 7 and created the stack Continuum B6/7 non-detections. We present the estimated fluxes and weighted galaxy properties in Table ~\ref{table_stacking}. We only obtain a significant detection (over $1.5\sigma$) for the Continuum B6/7 all arcs stack. We were able to obtain the dust masses for the continuum stacks as the stacked properties fulfill the requirement $\lambda_\mathrm{rest} > 250 \mu$m.

\begin{table*}
\caption{Stacking results for CO and continuum emissions, including intrinsic properties for the composite galaxy. }             
\label{table_stacking}      
\centering          
\begin{tabular}{l c c c c D{,}{\pm}{-1} D{,}{\pm}{-1} D{,}{\pm}{4.4} }     
\hline\hline       
Stacking& $\Delta \nu$ $S_\nu$&$S_\nu$ &$z$&  $M_{mol}$ 
& \multicolumn{1}{c}{Log(${M}_{*}$)}
&\multicolumn{1}{c}{Log(SFR)} \\ 
&[mJy $\cdot$ Km $\rm{s}^{-1}$]& [$\mu$Jy]&
&\multicolumn{1}{c}{[$10^8 \rm{M}_{\odot}$]} 
&\multicolumn{1}{c}{[$\rm{M}_{\odot}$]} 
&\multicolumn{1}{c}{[$\rm{M}_{\odot}\rm{yr}^{-1}$]} \\
\hline                    
CO all arcs&1.3$\pm$0.6&-&2.68 &2.0$\pm$0.9  & 9.6,0.09 & 0.96,0.07 \\  
CO some arcs&1.2$\pm$0.8&-& 2.69 &1.9$\pm$1.2 & 9.59,0.09 & 0.95,0.07\\
Continuum B3/4 all arcs&-&<1.2 & 2.74 &  <340    & 9.62,0.09 & 0.98,0.07\\
Continuum B6/7 all arcs&-& 4.9$\pm$0.9&2.96&  22$\pm$4 & 9.6, 0.1& 1.0,0.1\\
Continuum B6/7 some arcs&-& 4$\pm$2& 2.90 &12 $\pm$ 6 & 9.59,0.07 & 0.9,0.1\\
\hline                  
\end{tabular}
\tablefoot{Column 2 displays the integrated flux, Column 3 displays the flux, Column 4 displays the redshift, Column 5 displays the molecular gas mass, Column 6 displays the logarithm of stellar mass, and Column 7 displays the logarithm of the SFR.}
\end{table*}

\subsection{Molecular gas masses}
\subsubsection{Molecular gas mass using CO emissions}

As we are working with subsolar metallicities, we decided to employ an $\alpha_{\rm{CO}}$ model that accounts for the photo-dissociation of molecular gas in these types of galaxies \citep{genzelCombinedCODust2015}. Thus, we selected the first model (a) proposed in \citet{genzelCombinedCODust2015}. We also considered the \cite{accursoDerivingMultivariateACO2017} model as it was created for galaxies with similar metallicities \citep{accursoDerivingMultivariateACO2017}. However, as all the scaling relations we used to compare our data were derived using the \cite{tacconiPHIBSSUnifiedScaling2018} model, a similar model was considered more adequate.\\

\begin{figure}
\centering
\includegraphics[width=0.9\hsize]{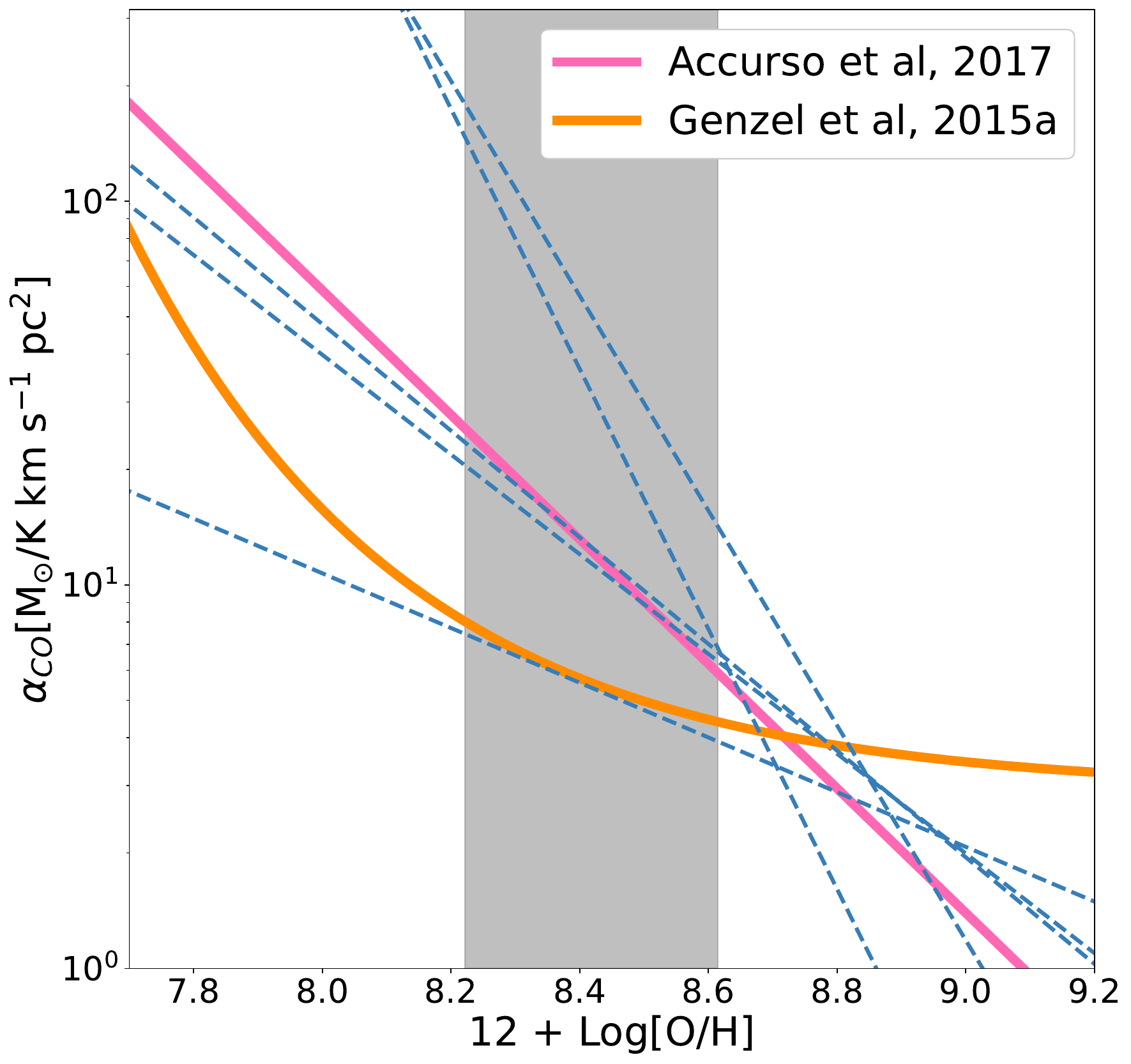}
   \caption{Comparison of different $\alpha_{\rm{CO}}$ models across varying metallicities. The shaded region indicates the metallicity range of our sample. Dashed lines show several models that were not considered for this work \citep{wilsonMetallicityDependenceCOtoH1995,magdisGOODSHerschelGastodustMass2011,genzelMetallicityDependenceCO2012,schrubaLowCOLuminosities2012,genzelCombinedCODust2015,accursoDerivingMultivariateACO2017,tacconiPHIBSSUnifiedScaling2018,maddenTracingTotalMolecular2020}.}
    \label{fig_alfa}
\end{figure}

We must use Equation~\ref{eq_alpha} to obtain the molecular gas mass, which only applies to CO(1-0). As we worked with other J transitions (J=3,4,5), an additional conversion factor (${r}_{\rm{Ji}}$) was required. We define ${r}_{\rm{J1}}$ in Equation ~\ref{defr}, where ${L'}_{\rm{CO(J-(J-1))}}$ and  ${L'}_{\rm{CO(1-0)}}$ are the CO luminosities in the transitions (J-(J-1)) and (1-0), respectively. For this study, we obtained relationships for ${r}_{\rm{51}}$, ${r}_{\rm{41}}$, and ${r}_{\rm{31}}$, which enable us to determine the CO(1-0) luminosity for each arc.\\

\begin{equation} \label{defr}
    {r}_{\rm{J1}} = \frac{{L'}_{\rm{CO(J-(J-1))}}}{{L'}_{\rm{CO(1-0)}}}
.\end{equation}

\begin{figure}
\centering
\includegraphics[width=0.9\hsize]{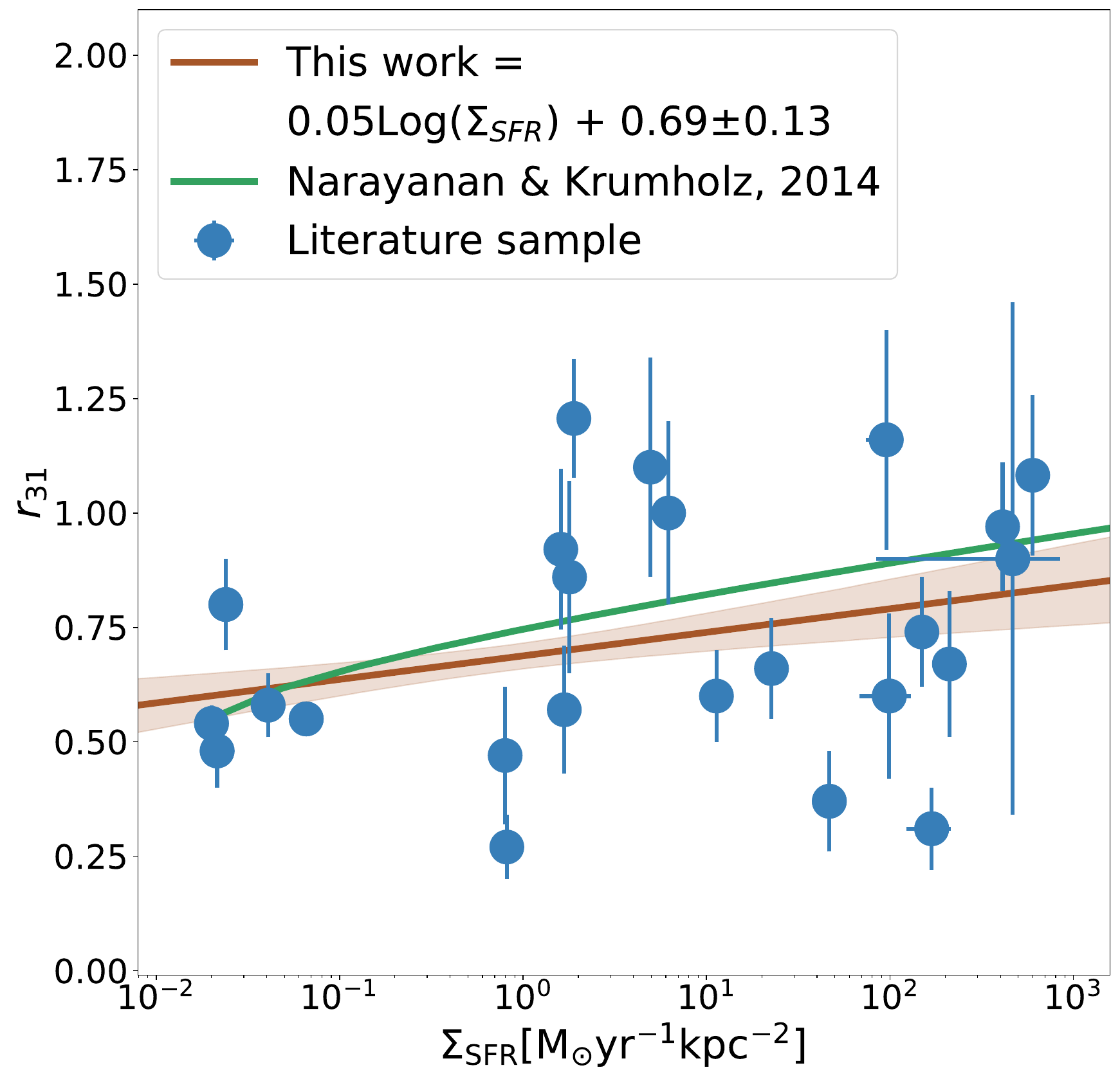}
   \caption{Relationship between $\Sigma_{\rm{SFR}}$ and ${r}_{\rm{31}}$. Blue dots denote data points used in constructing our model. The brown line represents our estimated model with a 1$\sigma$ error range, while the green line depicts the \cite{narayananTheoryExcitationCO2014} model for unresolved sources.}
    \label{R31}
\end{figure}

\begin{figure}
\centering
\includegraphics[width=0.9\hsize]{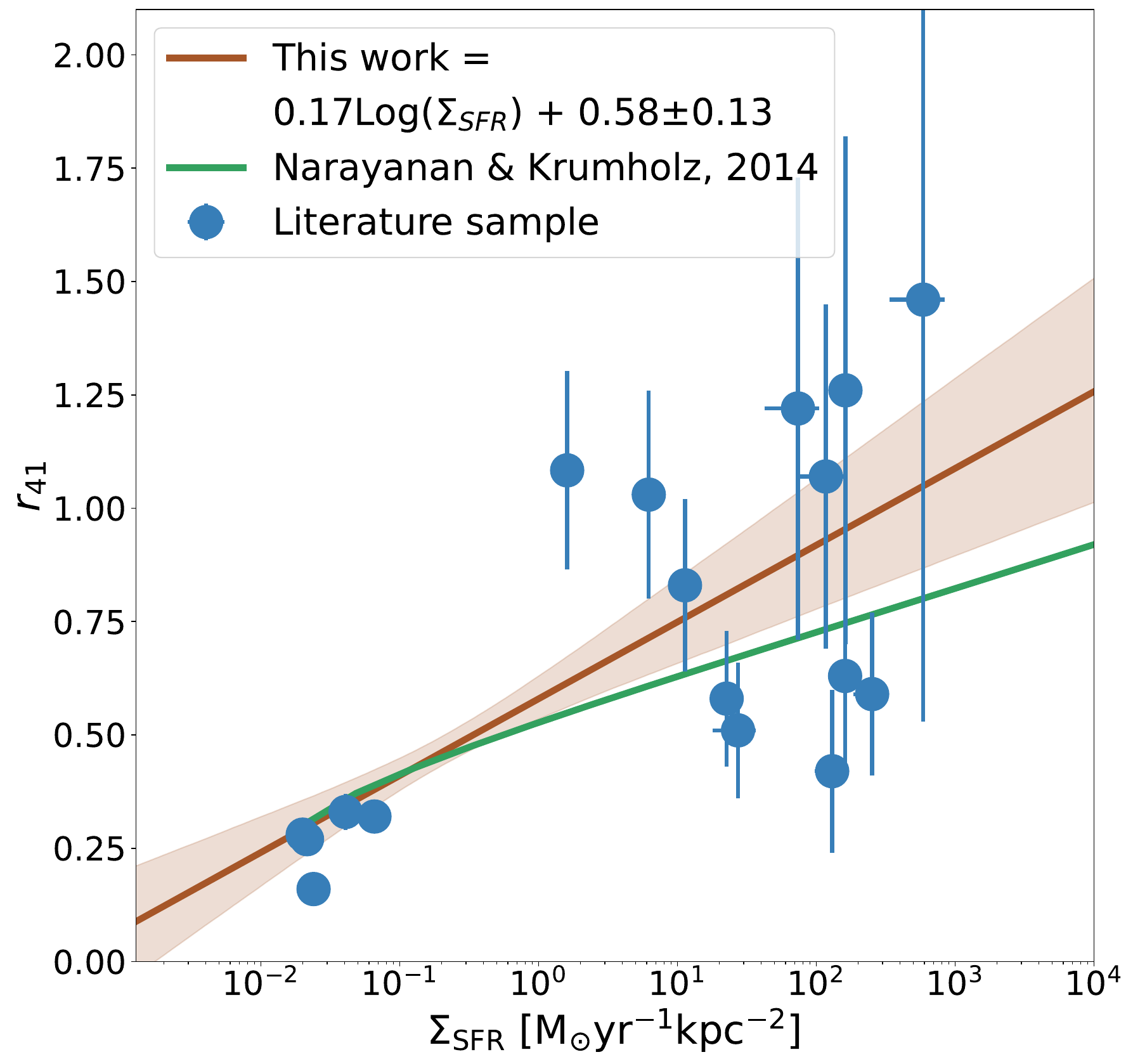}
   \caption{Same as Fig.~\ref{R31} but for ${r}_{\rm{41}}$. }
    \label{R41}
\end{figure}

Studies such as that of \citet{narayananTheoryExcitationCO2014} have proposed a relationship between $\Sigma_{\rm{SFR}}$ and ${r}_{\rm{J1}}$, as this parameter serves as a tracer of temperature and density, both of which influence the CO spectral line energy distribution (SLED)\citep{narayananTheoryExcitationCO2014}. In their study, these authors used models based on resolved and unresolved simulations across various values of J to obtain relationships between these quantities. These models are presented as green lines in Fig.~\ref{R31} and Fig.~\ref{R41}. As a way to obtain more realistic results, we decided to compile an observational dataset \citep{taniguchiWhatControlsStar1998,papadopoulosMolecularGasLuminous2012,riechersDustobscuredMassiveMaximumstarburst2013,daddiCOExcitationNormal2015,hatsukadeMolecularGasProperties2019,brisbinNeutralCarbonHighly2019,kaurJanskyVeryLarge2022,henriquez-brocalMolecularGasProperties2022,lenkicCOExcitationHighz2023,castilloVLALegacySurvey2023} for different ${r}_{\rm{J1}}$ values and their respective $\Sigma_{\rm{SFR}}$. This dataset can be found in Zenodo (See Sect.~\ref{data}). We fitted a model to the observational data sets for J= 3,4 and derived relations between both ${r}_{\rm{31}}$ and ${r}_{\rm{41}}$ and $\Sigma_{\rm{SFR}}$, which are illustrated in Fig.~\ref{R31} and Fig.~\ref{R41}, respectively. These relations were then used to estimate the corresponding ${r}_{\rm{J1}}$ values using the $\Sigma_{\rm{SFR}}$ obtained by the SED fits. As the observational models are consistent with the simulated models by \citet{narayananTheoryExcitationCO2014}, for the case of ${r}_{\rm{51}}$, where insufficient observational data were available, we opted to apply this simulated model. It is important to consider that both derived relations present significant scatter, which incorporates systematic errors into our molecular gas estimations.\\


 Finally, by using the obtained $\alpha_{\rm{CO}}$ and Equation ~\ref{eq_alpha}, we were able to determine the corresponding molecular gas masses. The magnification-corrected masses are reported in Table ~\ref{table_mol_mass}, along with the respective ${r}_{\rm{J1}}$ and $\alpha_{\rm{CO}}$ values.

\begin{table*}
\caption{Molecular gas masses derived from CO emission and dust continuum, corrected by magnification and accompanied by the specific parameters used for each arc.}             
\label{table_mol_mass}       
\centering          
\begin{tabular}{l c c c c c c}     
\hline\hline      
Arc
& $\alpha_{\rm{CO}}$& ${r}_{\rm{31}}$ & ${r}_{\rm{41}}$& ${r}_{\rm{51}}$& ${M}_{\rm{mol}}$       & ${M}_{mol}$ \\
& & & & & CO&(870 $\mu$\rm{m}/1.25\rm{mm}) \\

&[$ \rm{M}_{\odot}(\rm{Kkms}^{-1}\rm{pc}^{-2})^{-1}$]&&&&[$10^9 \rm{M}_{\odot}$]&[$10^9 \rm{M}_{\odot}$]\\ 
\hline                    
J0033    & 4.42$\pm$0.08   & -    & 0.60$\pm$0.5 & -    & 10$\pm$3          &  23$^{+7} _{-6}$ \\
J0108    & 5.8$\pm$0.3   & -    & 0.47$\pm$0.03 & -    & 30$\pm$17 &  <16              \\
J0900A   & 7.2$\pm$0.3   & -    & 0.56$\pm$0.05 & -    & <12                 &  <21              \\
J0900B   & 7.2$\pm$0.3   & -    & 0.59$\pm$0.6 & -    & <9                  &  <13              \\
J1050    & 7$\pm$1   & 0.69$\pm$0.03 & -    & -    & <1.6                &  7$\pm$2       \\
J1050-A  & 4.9$\pm$0.1   & 0.66$\pm$0.03 & -    & -    & 5$\pm$1           &  -               \\
J1226A   & 8.1$\pm$0.6   & -    & -    & 0.285$\pm$0.004 & <2.3               &  <1               \\
J1226B   & 8.1$\pm$0.6   & -    & -    & 0.231$\pm$0.003 & <4.2               &  8$\pm$2       \\
J1429    & 5.3$\pm$0.2   & 0.77$\pm$0.06 & -    & -    & <18                 &  <74              \\
J1527    & 5.9$\pm$0.4   & 0.74$\pm$0.04 & -    & -    & <10                 &  <9               \\
J2111    & 4.9$\pm$0.1   & 0.66$\pm$0.03 & -    & -    & <2.6                &  <7               \\
Sunburst & 5.5$\pm$0.2   & -    & 0.534$\pm$0.04 & -    & <0.4                &  < 1.2 \\
\hline                  
\end{tabular}
\tablefoot{Column 2 indicates the $\alpha_{\rm{CO}}$ conversion factor used for the calculation. Columns 3 to 5 show the ${r}_{\rm{J1}}$ values for J=3,4,5. Column 6 displays the molecular gas mass obtained using CO, and column 7 shows the molecular gas mass obtained using 870 $\mu$\rm{m}/1.25\rm{mm} dust continuum.}
\end{table*}

\subsubsection{Molecular gas mass obtained with dust mass}
We also derived the molecular gas mass by relating it to the dust mass using the gas-to-dust ratio ($\delta_{\rm{GDR}}$). This value is as uncertain as $\alpha_{\rm{CO}}$ and also depends on metallicity. The model used exhibits a power-law dependency on metallicity: $\delta_{\rm{GDR}} \propto {Z}^\gamma$ \citep{tacconiPHIBSSUnifiedScaling2018}. We adopted the calibration used in \citet{tacconiPHIBSSUnifiedScaling2018}, where $\gamma = 0.85$, and the normalization $\delta_{\rm{GDR}}({Z}_\odot) = 100$, as reported in \cite{draineDustMassesPAH2007}. We selected this relation as it should hold for the studied metallicity range \citep{remy-ruyerGastodustMassRatios2014, leroyCOH2CONVERSIONFACTOR2011}, and because it allows a direct comparison with the \cite{tacconiPHIBSSUnifiedScaling2018} dataset.
\begin{equation}
    \label{dust_to_mass}
    \delta_{\rm{GDR}} = 100 \left (  \frac{Z}{Z_\odot} \right )^{-0.85}
.\end{equation}
We were able to derive molecular gas masses for each galaxy using the metallicities reported in Table ~\ref{table_metallicities} and the dust masses obtained with Equation ~\ref{dust_to_mass}. Table ~\ref{table_mol_mass} presents the magnification-corrected values.

\subsubsection{Molecular gas comparison}

\begin{figure}
\centering
\includegraphics[width=1\hsize]{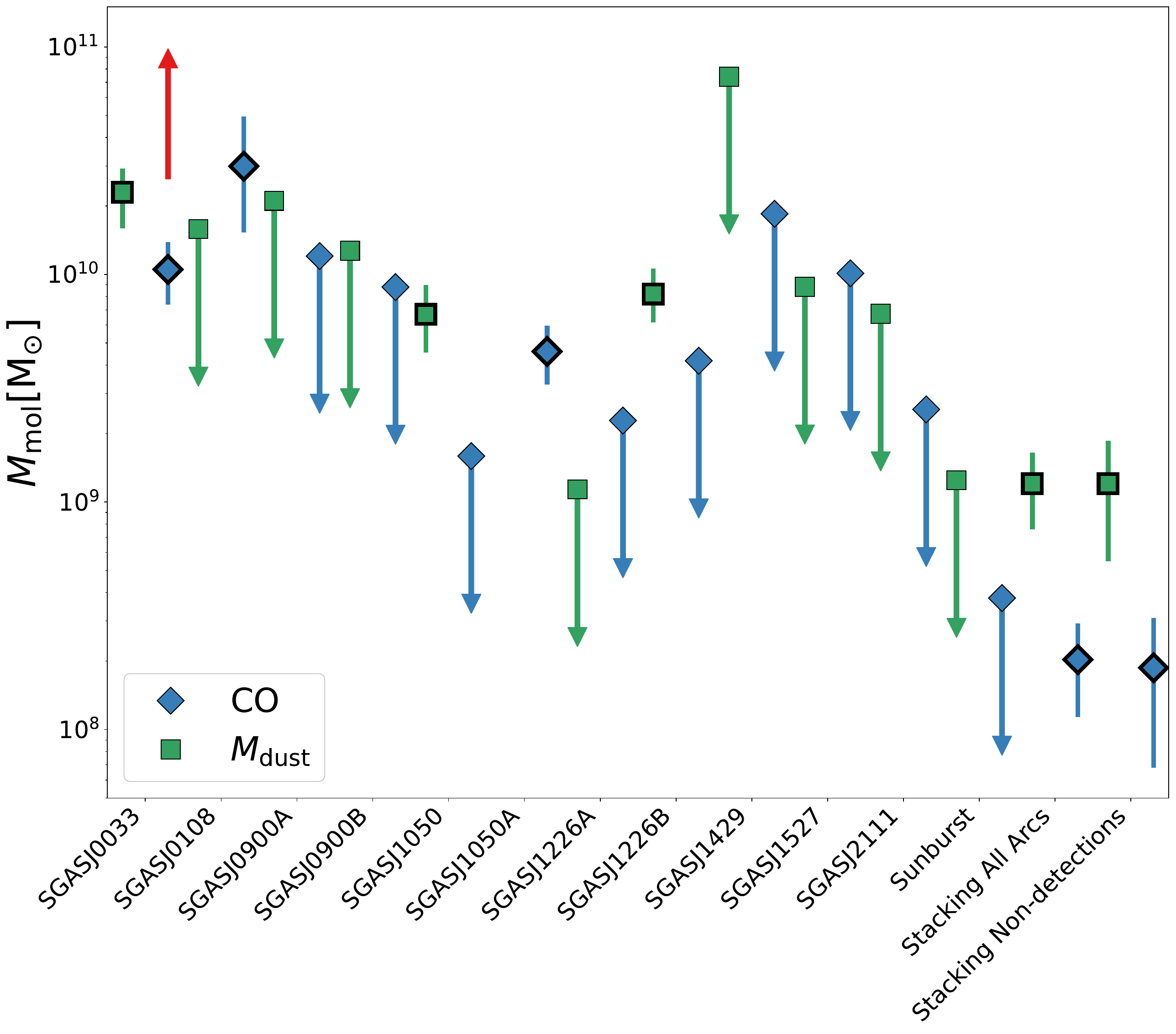}
   \caption{Comparison of the molecular gas mass estimates using the two methods discussed in the text. The red arrow shows by how much the CO points would be displaced if we had used the \cite{accursoDerivingMultivariateACO2017} model for $\alpha_{\rm{CO}} $. The stacking samples were created using weighted averages with weights of 1/$\sigma^2$. The stacking All Arcs considers all arcs while stacking non-detections only considers arcs without detection. Markers with a black edge and error bars indicate detections.}
      \label{comp}
\end{figure}

We determined the molecular gas masses using two distinct methods: CO emission and 870$\mu$m/1.25mm dust continuum. The corresponding results are presented in Fig.~\ref{comp}. In addition to using these methods, we measured the masses using the 2.1/3.1 \unit{mm} dust continuum, which only included upper limits. All values were at least one dex higher than the CO-obtained masses, and so they were not considered for this work. We find the most constraining method to be the estimation using CO emission. This CO-based approach is also generally consistent with the dust mass method at 870$\mu$m/1.25mm (bands 6 and 7). The red arrow in Fig.~\ref{comp} shows by how much the CO points
would be displaced if we had used the \cite{accursoDerivingMultivariateACO2017} model for $\alpha_{\rm{CO}} $ instead of that of \citet{genzelCombinedCODust2015}. In the figure, all detections are surrounded by a black edge.\\

For the remainder of the study, we primarily focus on CO-based molecular gas estimates, as they are known to only trace the molecular gas, while dust masses could possibly also trace atomic gas \citep{scovilleEvolutionInterstellarMedium2014}. Due to this, the CO-derived masses are expected to more accurately trace the molecular gas content.\\
 
\section{Discussion}\label{discussion}

\subsection{The galaxies in our sample in context}

We selected datasets from recent studies to contextualize our galaxies within the current scientific landscape. First, we used a sample of 137 high-redshift ($z>$2) SFGs from the PHIBSS catalog \citep{tacconiPHIBSSUnifiedScaling2018}, which are primarily located on the MS and span a stellar mass range of $\sim$ $10^{9.8}$-$10^{11.8}\,\rm{M}_\odot$. Second, we included the complete xCOLD GASS catalog \citep{saintongeXCOLDGASSComplete2017}, which consists of 306 galaxies at approximately $z\sim$0, and covers a stellar mass range of $\sim$ $10^9$ to $10^{11.3}\,\rm{M}_\odot$. These galaxies were selected to show the general behavior of SFGs at different masses and redshifts. 

We incorporate a sample of strongly lensed SFGs from four studies \citep{saintongeValidationEquilibriumModel2013, dessauges-zavadskyMolecularGasContent2015, mottaCosmicSeagullHighly2018,tsujitaALMALensingCluster2024a}. This sample covers a redshift range between 1.4 and 3.65 and Log($M_*$/M$_{\odot}$) between 9.26 and 11.49. All measurements of molecular gas masses were obtained using CO emission. To ensure consistency and to enable a robust comparison, we changed the $\alpha_{\rm{CO}}$ parameter used for the data to the \cite{genzelCombinedCODust2015} value. All these galaxies were analyzed using the same gas estimation method as that used in the present study and have characteristics similar to those of the studied galaxies, enabling a direct comparison with our data.

Additionally, we include galaxies identified as MS starbursts (MS SBs) by \citet{gomez-guijarroGOODSALMAStarburstsMain2022}. These galaxies were selected from the ALMA-GOODS 2.0 survey, which comprises low-resolution 1.1 mm observations that cover a redshift range between 1.314 and 4.73. MS SBs were identified based on a maximum deviation from the MS of $\Delta_{\rm{MS}}$ $\equiv$ SFR/$\rm{SFR}_{\rm{MS}}$  $\equiv$  3, and their short gas depletion timescales. Though dust masses were used  to obtain the molecular gas properties for  this sample, we consider it may provide valuable insight as it shows a similar behavior to our sample.

It is important to note that all these measurements represent detections rather than a complete census, which may introduce certain biases in our analysis. As galaxies with upper limits tend to be less bright, their removal might introduce a bias, skewing the results toward easier-to-detect galaxies. This effect is significant at high redshift, where the Malmquist bias is more significant. Due to this, the scaling relations used as a comparison might differ at higher redshifts and could not represent the complete population of the time. By adding a high-redshift gravitationally lensed sample that traces lower masses, we can slightly reduce this latter bias and achieve a more robust comparison.

\begin{figure}
\centering
\includegraphics[width=0.9\hsize]{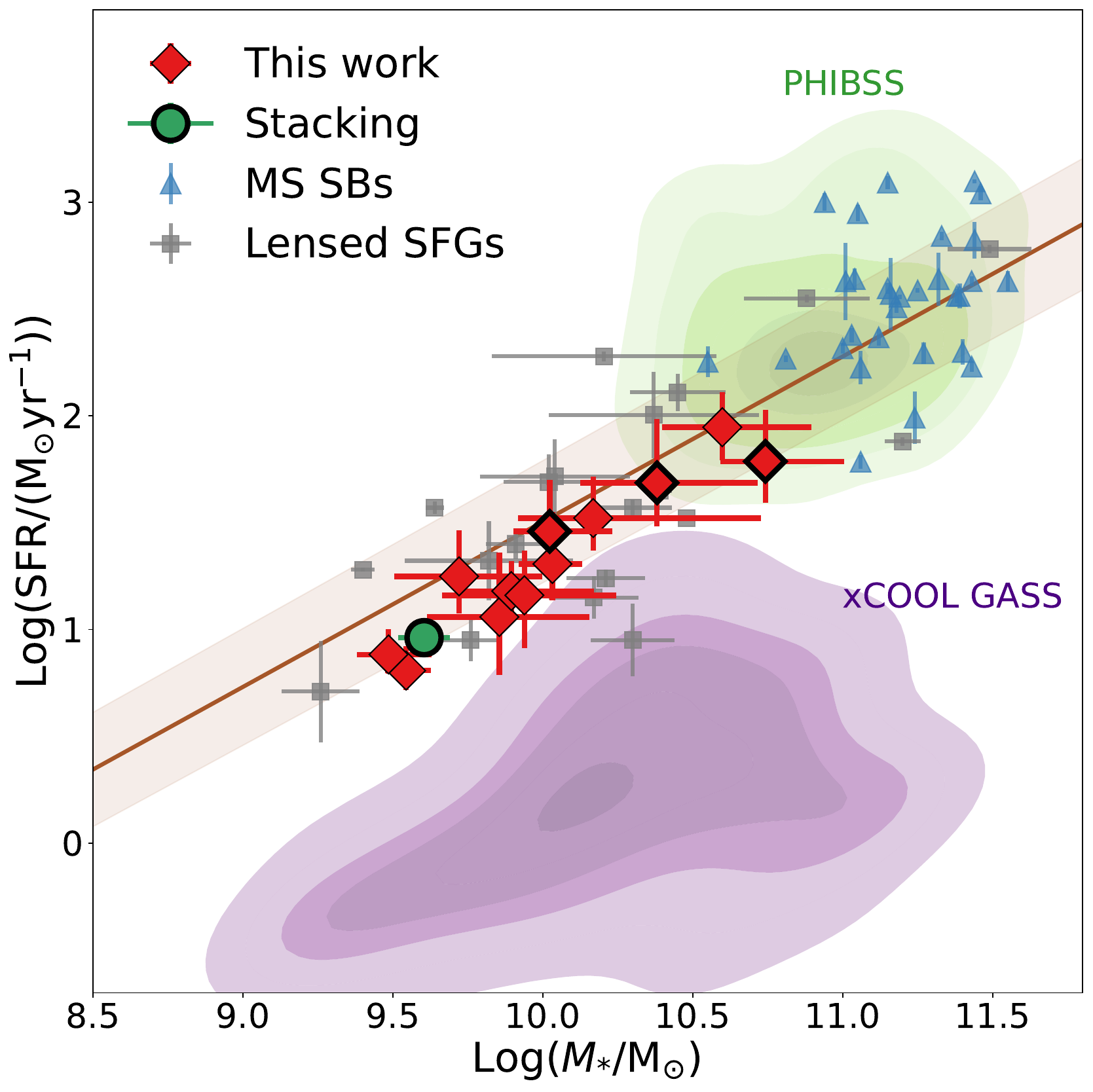}
   \caption{SFR as a function of stellar mass, known as the galaxy MS. The brown line denotes the MS at redshift 2.5 established by \citet{speagleHighlyConsistentFramework2014}. Gray markers represent lensed SFGs \citep{saintongeValidationEquilibriumModel2013,dessauges-zavadskyMolecularGasContent2015,mottaCosmicSeagullHighly2018,tsujitaALMALensingCluster2024a}, while blue markers are MS SBs from \citet{gomez-guijarroGOODSALMAStarburstsMain2022}. The purple contours are density plots of the xCOOL GASS data, while the green contours are the PHIBSS data from \cite{tacconiPHIBSSUnifiedScaling2018}. Red markers indicate the galaxies from this work; they represent detections when they have a black edge and error bars. The stacking value corresponds to the All Arcs value.}
      \label{MS}
\end{figure}

In Fig.~\ref{MS}, we present the data from our study alongside the recent literature values, comparing them to the MS relationship at redshift 2.5 derived from \citet{speagleHighlyConsistentFramework2014}. Notably, all our galaxies and stacking values fall within the MS, indicating that we are dealing with typical MS SFGs from the epoch.

\begin{figure}
\centering
\includegraphics[width=0.9\hsize]{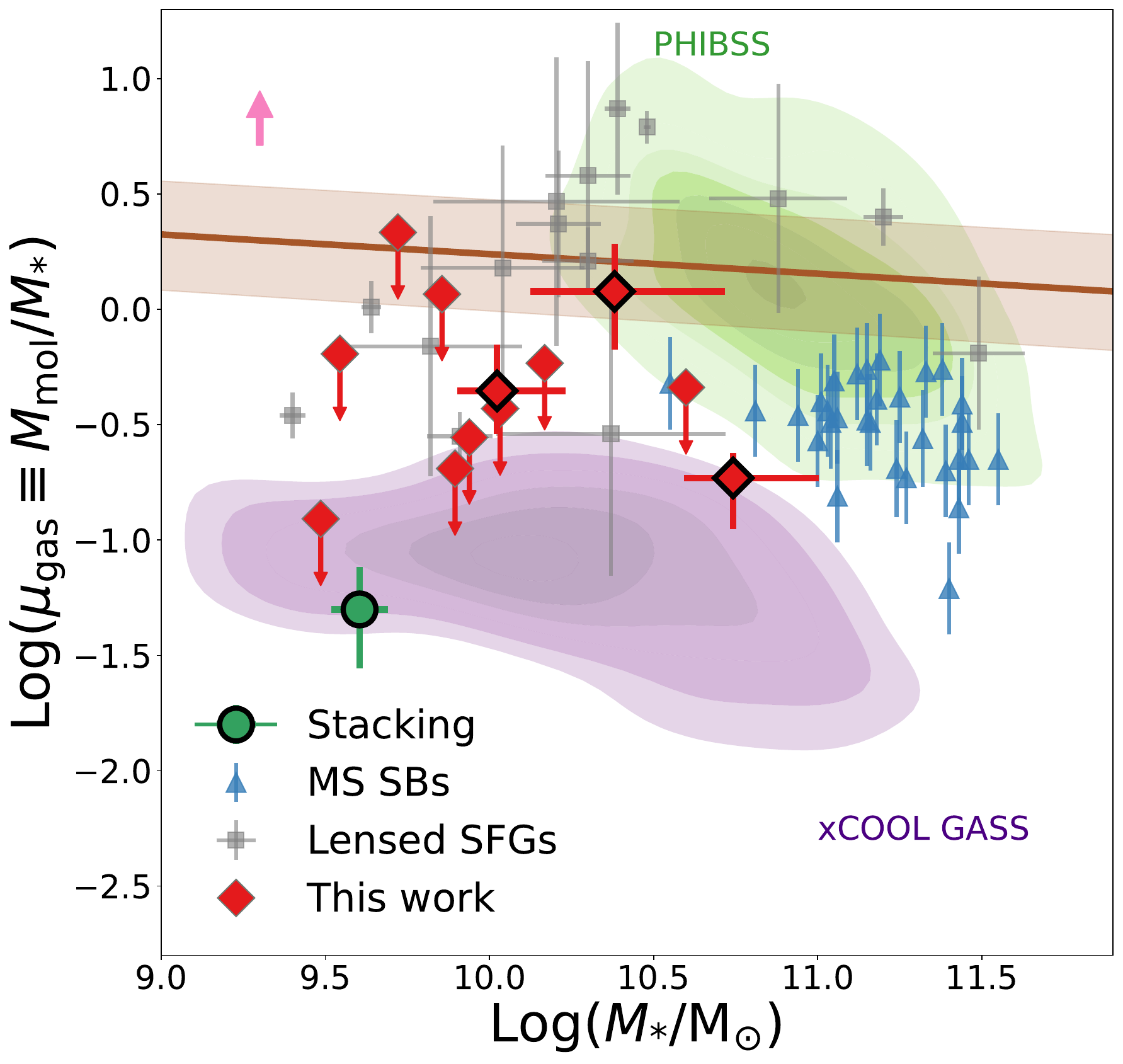}
   \caption{Same as Fig.~\ref{MS} but for the gas fraction as a function of stellar mass. The brown line denotes the scaling relation established by \citet{tacconiPHIBSSUnifiedScaling2018} for MS galaxies. The pink arrow corresponds to how much the CO points would move upwards if we had used the \cite{accursoDerivingMultivariateACO2017} model for $\alpha_{\rm{CO}} $.}
      \label{gas_frac}
\end{figure}

The ratio between molecular gas mass and stellar mass, known as the gas fraction ($\mu_{\rm{gas}}$), offers insights into the relative abundance of molecular gas compared to stellar mass within galaxies, that is, gas already converted into stars. In Fig.~\ref{gas_frac}, we depict how this ratio varies with stellar mass. The brown line represents the scaling relation for MS galaxies established by \citet{tacconiPHIBSSUnifiedScaling2018}. This model was chosen for comparison as it was seen to be the one that best follows the CO-based observations at the studied redshifts \citep{sandersCOEmissionMolecular2023}. Our analysis reveals that most of our galaxies align with or fall below the scaling relation, with some of those below this relation even representing upper limits, and thus not adhering to the relation. Notably, the gas fractions observed in our study are comparable to those reported in \citet{gomez-guijarroGOODSALMAStarburstsMain2022}, although the galaxies we study  have significantly lower stellar masses. While some lensed sources \citep{saintongeValidationEquilibriumModel2013, dessauges-zavadskyMolecularGasContent2015, mottaCosmicSeagullHighly2018,tsujitaALMALensingCluster2024a} agree with our targets, others lie on or above the scaling relation. Most of the sources with similar gas fractions to our sample have similar masses, which could indicate that this is a normal behavior at lower masses. Additionally, galaxies derived from our stacking analysis appear to fall below the relationship. As the stacking represents a low-mass galaxy with low metallicity, this suggests that lower-mass lensed galaxies may exhibit a lower molecular gas fraction compared to their higher-mass counterparts\\

\begin{figure}
\centering
\includegraphics[width=0.9\hsize]{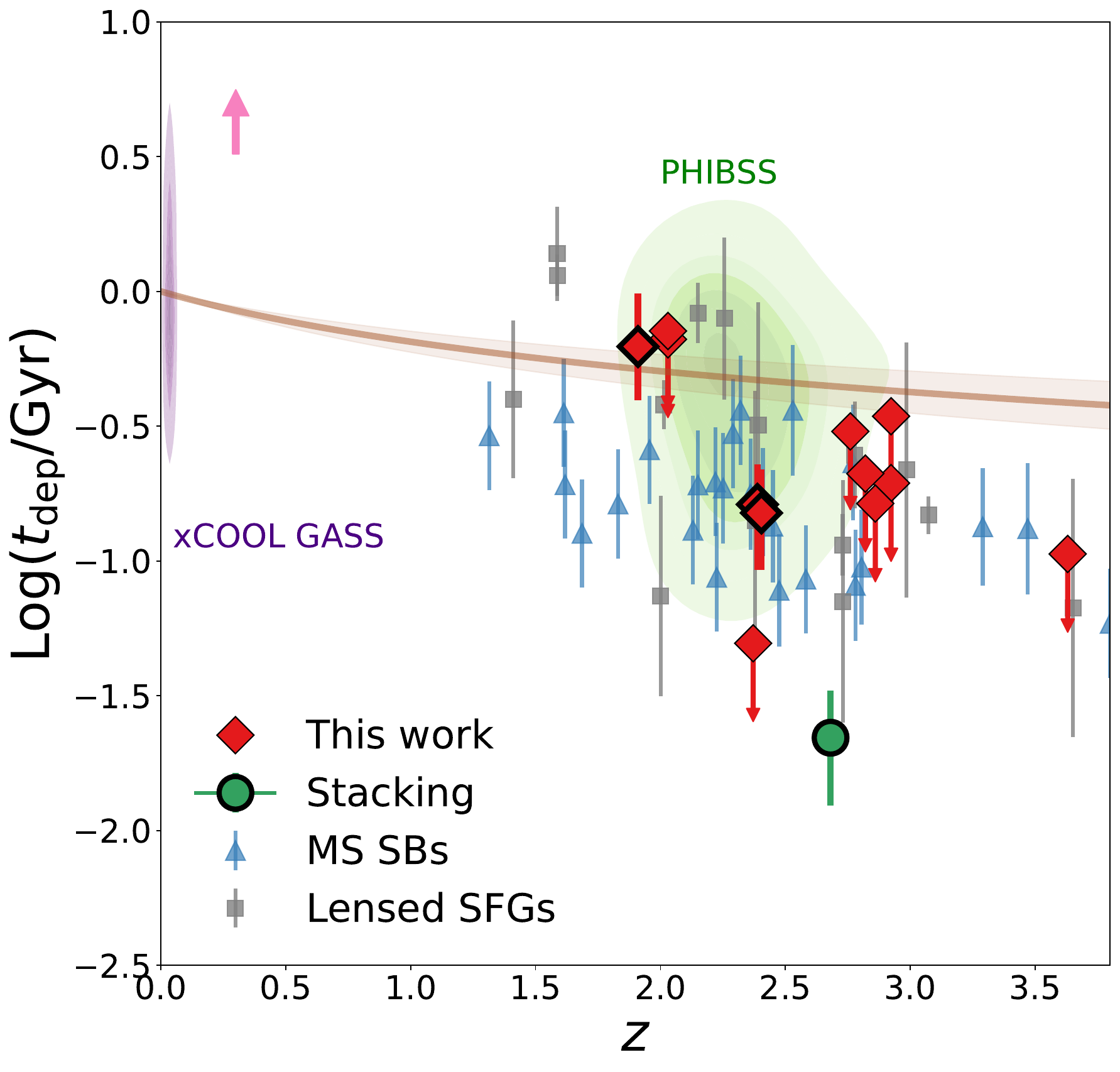}
   \caption{Same as Fig.~\ref{MS} but for depletion time as a function of redshift. The brown line represents the scaling relation established by \citet{tacconiPHIBSSUnifiedScaling2018}. The pink arrow corresponds to how much the CO points would move upwards if we were to use the \cite{accursoDerivingMultivariateACO2017} model for $\alpha_{\rm{CO}} $.}
      \label{t_dep}
\end{figure}

The depletion time, defined as the ratio of molecular gas mass (${M}_{\rm{mol}}$) to SFR, provides insight into how long a galaxy can sustain its current level of star-formation activity in the absence of new material. In Fig.~\ref{t_dep}, we illustrate how this property varies with redshift, with the brown line representing the scaling relationship proposed by \citet{tacconiPHIBSSUnifiedScaling2018}. This model was chosen as it was seen to be the one that best follows the CO-based observations at the studied redshifts \citep{sandersCOEmissionMolecular2023}. Our analysis reveals that most of our data points fall below this relationship, with some upper limits extending beyond it. This suggests that the targeted galaxies are experiencing accelerated depletion of their molecular gas reservoirs and are likely running out of fuel more rapidly than typical galaxies at similar redshifts. These findings are consistent with those reported in \citet{gomez-guijarroGOODSALMAStarburstsMain2022} and with what is observed in some lensed galaxies. However, other lensed galaxies \citep{saintongeValidationEquilibriumModel2013,dessauges-zavadskyMolecularGasContent2015,mottaCosmicSeagullHighly2018,tsujitaALMALensingCluster2024a} appear to be above this relationship. Furthermore, the data points from our stacking analysis also fall below the scaling relation, indicating that, on average, our galaxies exhibit relatively short depletion times. This suggests that the studied galaxies as a whole are undergoing rapid consumption of their molecular gas reservoirs.\\
\subsection{Interpretation and possible caveats}

The low depletion times and gas fractions observed in the studied galaxies may indicate a molecular gas deficit within these systems, given that the SFRs are consistent with being in the star-forming MS. This characterization aligns with the findings reported by \citet{gomez-guijarroGOODSALMAStarburstsMain2022}, who identified similar properties in galaxies they classified as MS SBs; despite being located on the MS, these galaxies exhibit low depletion times and gas fractions. In this study, we propose several possibilities to explain the observed deficit and discuss the validity of each scenario.

One possibility to explain the apparent gas deficit is the significant systematic errors in estimating the molecular gas masses. As shown in 
Fig.~\ref{fig_alfa}, there are several recipes for estimating $\alpha_{\rm{CO}}$. We can observe that the \citet{genzelCombinedCODust2015} model adopted in our study provides one of the lowest values for $\alpha_{\rm{CO}}$ at the studied metallicities. If we were to use other models that also account for photo-dissociation, such as that of \citet{maddenTracingTotalMolecular2020}, the $\alpha_{\rm{CO}}$ values would be larger, resulting in higher molecular gas mass estimates. Thus, the gas fractions would be higher by approximately 1 dex. However, the \citet{maddenTracingTotalMolecular2020} model was not chosen as its values were suggested by  \citet{ramambasonModelingMolecularGas2024} to be upper limits rather than realistic estimates. Another model considered was \cite{accursoDerivingMultivariateACO2017}, which would cause the gas deficit to disappear. However, considering that the scaling relationships were created using $\alpha_{\rm{CO}}$ values similar to \cite{genzelCombinedCODust2015}, if we were to re-estimate the scaling relations to use the \cite{accursoDerivingMultivariateACO2017} values, there is a probability that the deficit may still appear. Another important error is the one introduced by the scatter of the obtained relationships for ${r}_{\rm{41}}$ and ${r}_{\rm{31}}$. Though this error was estimated and propagated throughout this work, it may have affected the results.

We also considered the possibility that lower-mass galaxies have a more extensive atomic gas reservoir relative to their molecular gas reservoir. In both our study and those used for comparison, the gas fraction was approximated from $\mu_{\rm{gas}} = ({M}_{\rm{mol}}+{M}_{\rm{atomic}})/{M}_*$ to $\mu_{\rm{gas}} \sim {M}_{\rm{mol}}/{M}_*$, as this is considered valid at $z>0.4$ \citep{tacconiPHIBSSUnifiedScaling2018}. It is essential to consider that with increasing mass, the fraction of molecular gas compared to atomic gas increases, meaning lower-mass galaxies should contain a higher fraction of atomic gas \citep{poppingEvolutionAtomicMolecular2014,lagosMolecularHydrogenAbundances2015}. As \citet{tacconiPHIBSSUnifiedScaling2018} only studies high-mass galaxies at high redshifts, the simplification is valid for their study, and the scaling relations should not vary if we consider the complete definition of $\mu_{\rm{gas}}$. Interestingly, \cite{messiasHIContentCosmic2024} found there might be evidence for an increasing atomic gas reservoir between the redshifts $z\sim$1.5 and $z\sim$2.5, which could indicate that the  galaxies at cosmic noon have higher atomic gas reservoirs. As our galaxies are low-mass galaxies at cosmic noon, a relatively higher atomic gas content could be a suitable explanation for the observed deficit. A way of estimating the atomic gas mass would be with the molecular gas mass obtained using the dust observations, as initially \cite{scovilleEvolutionInterstellarMedium2014} considered the gas obtained with dust masses to be composed of both atomic and molecular gas. Due to this, we decided to use the 870 $\mu$\rm{m}/1.25\rm{mm} dust continuum and CO molecular gas masses of the All galaxies stacks and compare them. From this, we obtained that the molecular mass estimated with CO was only 9\% of the molecular mass obtained with the dust mass, which means that possibly 91\% of the gas is in atomic form.

Another possible explanation for the apparent observed gas deficit is the presence of what is known as CO-dark gas. Reduced dust abundance permits UV radiation to photo-dissociate CO molecules more effectively at low metallicities, such as in the galaxies under study. Consequently, while $\rm{H}_2$ remains shielded from photo-dissociation, we are left with the same amount of molecular gas but with only a central CO core and a \ci \ cloud \citep{maddenTracingTotalMolecular2020}. This scenario implies that CO may not serve as an effective tracer for the entire molecular gas content, potentially capturing less than 30\% of the gas present in the galaxy. Although efforts were made to mitigate this effect by incorporating an $\alpha_{\rm{CO}}$ value that accounts for photo-dissociation, this adjustment may still introduce a systematic bias. As a possible solution, works such as those of \citet{maddenTracingTotalMolecular2020} and \cite{heintzDirectMeasurementLuminosity2020} suggested the use of \ci and \cii as tracers of the molecular gas in low-metallicity regions. However, both tracers bring with them possible difficulties. At high redshift and low metallicity, the \ci emission of a galaxy tends to be significantly lower than that of \cii as it requires a higher density to produce collisional excitations \citep{maddenTracingTotalMolecular2020}. This means that \cii emission is easier to detect in these conditions. However, it has been suggested that this line might also trace atomic gas, which could make it a less robust tracer \citep{maddenTracingTotalMolecular2020}.
To estimate whether or not CO-dark gas is present in our sample, we obtained the molecular gas masses using \cii \ for the arcs J1226A and J1226B. The unlensed values were  $(4.9\pm1.1)\times10^{8}$ M$_{\odot}$  and $(1.2\pm 0.2)\times10^{10}$ M$_{\odot}$, respectively. These results will be presented in  Solimano et al. (in prep.). The estimation details are in Appendix ~\ref{sec:cplus}. In the case of J1226B, this value is lower than the upper limit obtained by CO, indicating consistency between both methods. However, for J1226B, the value is significantly higher, supporting the idea that CO may not be a good tracer of molecular gas in some cases. Additionally, from what we obtain from the stacking, the CO only traces 9\% of the gas traced by the dust, meaning part of the galaxies' gas could be CO-dark.

Finally, a plausible explanation for the apparent gas deficit observed in our study is that it is accurate. In this scenario, we can draw parallels with the MS SBs proposed by \citet{gomez-guijarroGOODSALMAStarburstsMain2022}. While these authors primarily identified high-mass galaxies as MS SBs (Log($M_*$)>10.55), our results suggest that low-mass galaxies could exhibit similar characteristics, namely low depletion times and low gas fractions, while remaining in the MS. Though our depletion times and gas fractions are larger than those of these latter authors, most are upper limits, meaning they could eventually coincide. \citet{gomez-guijarroGOODSALMAStarburstsMain2022} proposed two potential mechanisms by which galaxies could achieve these characteristics. The first scenario involves the compaction of the galaxy's core following a gas-rich merger, leading to gas funneling toward the center. This process increases the SFR, causing the galaxy to move upward in the MS plot. Subsequently, as the gas is consumed, the galaxy gradually lowers its SFR, traversing through the MS phase. The second scenario states that gas is funneled toward the galaxy's center, increasing SF activity. As the gas supply is depleted over time, the star-forming region becomes more compact, sustaining a high SFR. Eventually, this process may result in forming a compact star-forming core, which continues to fuel SF activity. Of the two, this second scenario is the most probable explanation for the observed characteristics in our galaxies. Unlike mergers, which typically result in the formation of more giant galaxies, the sustained SF with less gas than expected suggests a mechanism wherein gas is efficiently funneled toward the galaxy's center, enabling sustained SF despite the gas deficit.

After considering various explanations for the observed gas deficit in the studied galaxies, we find that a combination of factors is likely at play. Our data show that the most likely scenario is a mixture of CO-dark gas and a relatively high atomic gas reservoir. From the stacked values, we obtain that CO only accounts for 9\% of the gas traced by the dust emission, which could indicate that the rest of the gas is in atomic form. However, due to the low metallicity of the studied galaxies, it is paramount to consider that some of the gas might not be well traced by CO. Due to this, it is more accurate to consider the 91\% of atomic gas to be an upper limit of atomic gas rather than the absolute value.

\section{Summary and conclusions}\label{conclusions}

This research focuses on 12 gravitationally lensed low-mass SFGs with redshifts between 1.9 and 3.6. We used ACA observations to search for CO emission lines and analyze the galaxies' dust continuum to estimate different molecular gas mass values. We obtained two different values, one using CO and another using dust mass. Additionally, we used supplementary HST and JWST data for SED fitting, allowing us to obtain the intrinsic properties of the galaxies after correcting by the gravitational magnification factor. 

CO was only detected in 3 of the 12 arcs (J0033, J0108, and J1050A). In contrast, 3 galaxies were detected in 870 $\mu$\rm{m}/1.25\rm{mm} dust continuum (J0033, J1050, J1226B), with no galaxies detected in 2.1/3.1 \unit{mm} dust continuum. Additionally, we created stacked versions of our data to gain insights into the average properties of our sample. 

When comparing our data with reported scaling relations \citep{speagleHighlyConsistentFramework2014,
tacconiPHIBSSUnifiedScaling2018}, we find that all the galaxies belong in the MS. Some galaxies exhibit an apparent deficit of molecular gas compared with the extrapolation of scaling relations fitted to more massive galaxies at similar redshifts. To explain this deficit, we discuss four scenarios. We explored the possibilities of systematic errors associated with the molecular gas estimate and the scenario of a MS SBs. However, assuming our gas estimates are correct, it seems a combination of CO-dark gas in these low-metallicity galaxies and the predominant presence of gas in the atomic phase primarily drives the low molecular gas fraction and depletion times we observe.

This assumption arises from our stacked sample, as we obtain that 91\% of the gas may be in atomic form. However, we consider this an upper limit, as CO-dark gas may also affect the obtained CO measurements, which is corroborated by the \cii observations for J1226B. To further test this hypothesis, we must investigate the \ci and \cii emission for the studied galaxies, wherever possible. This would tell us whether or not our results are influenced by the presence of CO-dark gas and would allow us to determine whether the 91\% value obtained is an upper limit or an accurate estimate of the galaxy's atomic gas content. 
\section{Data Availability}\label{data}
Additional data, such as the observational dataset used for creating the $r_{J1}$ relations, can be found in Zenodo at the link \url{https://zenodo.org/records/14045715}.
\bibliographystyle{aa_url.bst}
\bibliography{catan_et_al_edited}

\newpage
\begin{appendix} 
 \onecolumn

\section{Gas mass from [C II] emission in SGASJ1226}\label{sec:cplus}
As suggested by \citet{maddenTracingTotalMolecular2020} and others, low metallicity molecular clouds have their CO emission heavily suppressed and are photon-dominated very deep into their inner regions. For this reason, several authors argue that in these regimes, the bulk of molecular gas is better traced by fine structure \cii emission line at a rest frequency of 1900.536 GHz (\cite{accursoDerivingMultivariateACO2017}, \cite{hughesVALESIIPhysical2017}, \cite{zanellaIIEmissionMolecular2018}, \cite{vizganInvestigatingIIToH2022}). To test this idea, we have obtained \cii observations of the J1226 field, covering two galaxies of our sample (ALMA program 2021.1.01337.S, PI: Solimano). A full analysis of these data will be presented in Solimano et al. (in prep.). The observations combine data from the 7m array and the 12m arrays. We reduced the visibility data using the standard pipeline version 6.2.1.7. Then, we synthesized datacubes using the multiscale deconvolver, adopting a channel width of 9.6\kms and restoring with a common circular beam size of 0.72 arcseconds. We achieved a point-source sensitivity of $1\sigma=2.4$ mJy beam$^{-1}$ channel$^{-1}$.

We visually inspected the datacube and extracted spectra from apertures that roughly match the apertures of arcs A.1 and B used for photometric extraction (see Fig.~\ref{fig_arcos}), but accounting for the larger beam compared to the HST and JWST imaging. The line was robustly detected in both arcs, with an integrated S/N of $\sim12$ in the brightest region (within arc B). We fitted Gaussians to the PB-corrected 1D spectra, with one kinematic component for arc A.1 and two for arc B. We measured total integrated fluxes of $3.3\pm0.5\,\mathrm{Jy}\kms$ and $15.72\pm0.24\,\mathrm{Jy}\kms$ in arcs A.1 and B, respectively. These fluxes translate into image-plane luminosities of $\mu L_{[\mathrm{C}\,\mathrm{II}]}^\mathrm{A.1} = (1.0 \pm 0.14)\times 10^9$L$_{\odot}$~and $\mu L_{[\mathrm{C}\,\mathrm{II}]}^\mathrm{B} = (4.81 \pm 0.07)\times 10^9$L$_{\odot}$.

Following \cite{zanellaIIEmissionMolecular2018}, we adopted a \cii-to-H$_2$ conversion factor of $\alpha_{[\mathrm{C}\,\mathrm{II}]}=30$ M$_{\odot}$L$_{\odot} ^{-1}$,which yields delensed molecular gas masses of $(4.9\pm1.1)\times10^{8}$ M$_{\odot}$ and $(1.2\pm 0.2)\times10^{10}$ M$_{\odot}$, for arcs A.1 and B, respectively. At face value, the inferred mass for SGASJ1226-A is consistent with our CO and dust upper limits, confirming a low gas fraction of $0.05\pm0.02$. Meanwhile, the gas mass inferred for SGASJ1226-B is significantly larger than our CO-based upper limit, albeit consistent with our value derived from the dust detection ($M_\mathrm{mol}=(9\pm 2)\times 10^{9}$ M$_{\odot}$). The \cii-derived mass implies a gas fraction as high as $1.2\pm0.3$.

These results suggest that despite the similar masses and colors of J1226-A and J1226-B, they seem to differ significantly in their cold ISM properties. On the one hand, galaxy B harbors a larger dust content, although not enough to shield CO molecules from photo-dissociation. On the other hand, galaxy A appears to be genuinely deficient in gas. We caution, however, that the $\alpha_{[\mathrm{C}\,\mathrm{II}]}$ factor might be as uncertain as $\alpha_\mathrm{CO}$ due to the unknown contribution of the atomic and ionized phases to the \cii luminosity, and the general lack of robust calibrations for large samples of galaxies. While some authors quote $\alpha_{[\mathrm{C}\,\mathrm{II}]}=10-18$ calibrated on simulated high-$z$ galaxies \citep{pallottiniZoomingInternalStructure2017, vizganInvestigatingIIToH2022}, others quote values as high as $\alpha_{[\mathrm{C}\,\mathrm{II}]}=130$ based on an empirical calibration of local dwarf galaxies \citep{maddenTracingTotalMolecular2020}.

\begin{figure}[!htb]
    \centering
    \resizebox{\hsize}{!}{\includegraphics{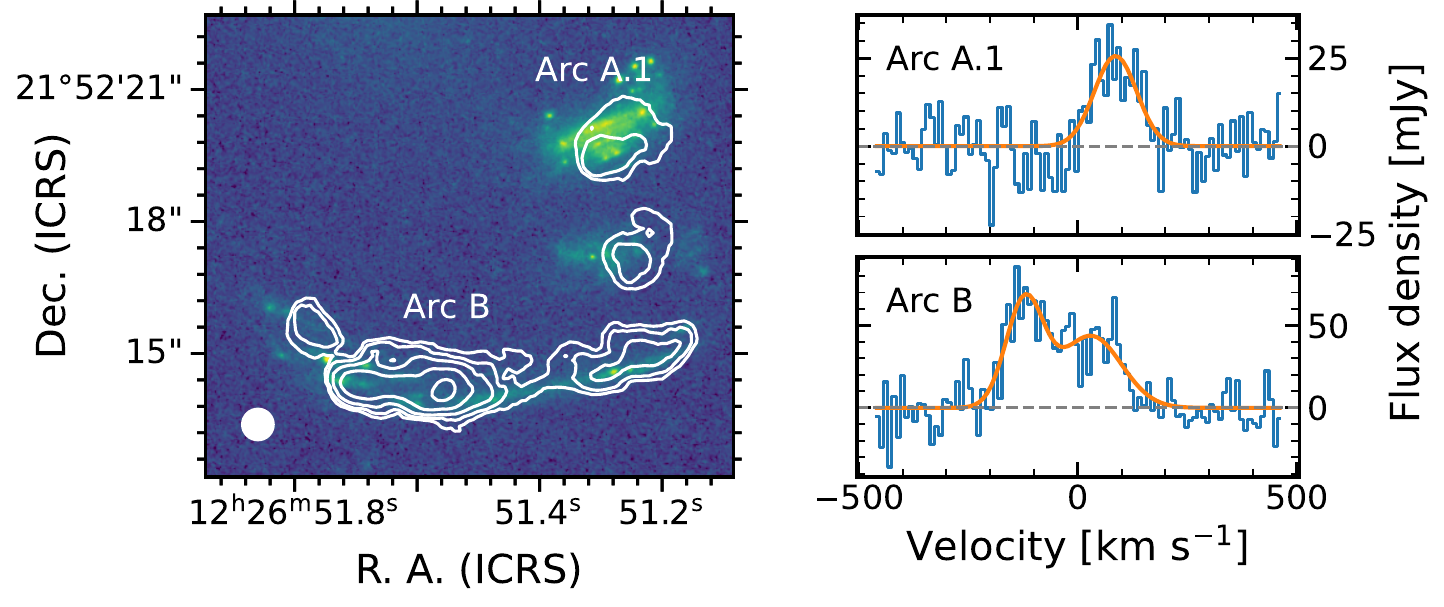}}
    \caption{HST ACS/F606W image of the J1226 field with ALMA \cii contours (white). The first contour is at the surface brightness level of 0.21 Jy \kms beam$^{-1}$, corresponding to S/N$\sim 3$. Subsequent levels increase as powers of $\sqrt{3}$. Inset axes show the integrated spectra of J1226-A.1 (top) and J1226-B (bottom).}
    \label{fig:cii-contours}
\end{figure}

\end{appendix}
\
\end{document}